\documentclass[a4paper,10pt]{article}

\usepackage{amsmath,amssymb,mathtools}     %%一些数学符号
\usepackage{color}
\usepackage{graphicx}
\usepackage{subfigure}
\usepackage{cite}                %%是引用格式变漂亮  %%PRD等APS模板必须禁用该项
\usepackage{hyperref}            %%超链接必须
\usepackage{multirow,makecell}   %%表格
\usepackage{textcomp}
\usepackage{wasysym}
\numberwithin{equation}{section}   %%公式按节编号
\def \be {\begin{equation}}
\def \ee {\end{equation}}
\def \ba {\begin{array}}
\def \ea {\end{array}}
\def \bea{\begin{eqnarray}}
\def \eea{\end{eqnarray}}
\def \nn {\nonumber}

\def \a {\alpha}
\def \b {\beta}

\def \G {\Gamma}
\def \d {\delta}

\def \L {\Lambda}
\def \s {\sigma}

\def \r {\rho}

\def \mA {\mathcal A}
\def \mB {\mathcal B}

\def \mD {\mathcal D}
\def \mE {\mathcal E}
\def \mF {\mathcal F}
\def \mG {\mathcal G}
\def \mH {\mathcal H}
\def \mI {\mathcal I}
\def \mJ {\mathcal J}
\def \mK {\mathcal K}
\def \mL {\mathcal L}

\def \mO {\mathcal O}
\def \mP {\mathcal P}
\def \mQ {\mathcal Q}
\def \mR {\mathcal R}

\def \mU {\mathcal U}
\def \mV {\mathcal V}

\def \mX {\mathcal X}

\def \mZ {\mathcal Z}

\def \p {\partial}
\def \f {\frac}

\def \mc {\mathcal}

\def \lt {\left}
\def \rt {\right}

\def \td {\tilde}

\def \inf {\infty}

\def \lag {\langle}
\def \rag {\rangle}

\def \ii {\mathrm{i}}

\def \tr {\textrm{tr}}

\def \and {{\textrm{and}}}

\def \CFT {{\textrm{CFT}}}
\def \T {{\textrm{T}}}
\def \spin {{\textrm{spin-}}}

\def \I {{\resizebox{0.2em}{\height}{\textrm{I}}}}
\def \II {{\resizebox{0.3em}{\height}{\textrm{\bf II}}}}
\def \III {{\resizebox{0.4em}{\height}{\textrm{\bf III}}}}
\def \IV {{\resizebox{0.4em}{\height}{\textrm{\bf IV}}}}
\def \V {{\resizebox{0.4em}{\height}{\textrm{V}}}}
\def \VI {{\resizebox{0.4em}{\height}{\textrm{\bf VI}}}}

\def \sI {{\resizebox{0.2em}{0.5em}{\textrm{I}}}}
\def \sII {{\resizebox{0.3em}{0.5em}{\textrm{\bf II}}}}
\def \sIII {{\resizebox{0.4em}{0.5em}{\textrm{\bf III}}}}
\def \sIV {{\resizebox{0.4em}{0.5em}{\textrm{\bf IV}}}}
\def \sV {{\resizebox{0.4em}{0.5em}{\textrm{V}}}}
\def \sVI {{\resizebox{0.4em}{0.5em}{\textrm{\bf VI}}}}

\def \oloop {{\textrm{1-loop}}}

\def \Area {{\textrm{Area}}}

\begin{document}

\title{\textbf{On one-loop entanglement entropy of two short intervals from OPE of twist operators}}
\author{Zhibin Li\footnote{lizb@ihep.ac.cn}~ and Jia-ju Zhang\footnote{jjzhang@ihep.ac.cn}}
\date{}

\maketitle

\vspace{-10mm}

\begin{center}
{\it
 Theoretical Physics Division, Institute of High Energy Physics, Chinese Academy of Sciences,\\
19B Yuquan Rd, Beijing 100049, P.~R.~China\\ \vspace{1mm}
Theoretical Physics Center for Science Facilities, Chinese Academy of Sciences,\\19B Yuquan Rd, Beijing 100049, P.~R.~China
}
\vspace{10mm}
\end{center}

\begin{abstract}
  We investigate the one-loop entanglement entropy of two short intervals with small cross ratio $x$ on a complex plane in two-dimensional conformal field theory (CFT) using operator product expansion of twist operators. We focus on the one-loop entanglement entropy instead of the general order $n$ R\'enyi entropy, and this makes the calculation much easier. We consider the contributions of stress tensor to order $x^{10}$, contributions of $W_3$ operator to order $x^{12}$, and contributions of $W_4$ operator to order $x^{14}$. The CFT results agree with the ones in gravity.
\end{abstract}

\baselineskip 18pt
\thispagestyle{empty}
\newpage

\tableofcontents

\section{Introduction}

Entanglement entropy plays an important role in characterizing the correlations of different parts in a many-body system \cite{nielsen2010quantum,petz2008quantum}. The usual way of calculating the entanglement entropy is the replica trick \cite{Callan:1994py,Holzhey:1994we}, in which one firstly calculates the general order $n$ R\'enyi entropy and then takes the $n \to 1$ limit.
It is usually not easy to calculate the entanglement entropy in a quantum field theory, but for a CFT (conformal field theory) that has a gravity dual in AdS (anti-de Sitter) background one can use the AdS/CFT correspondence \cite{Maldacena:1997re,Gubser:1998bc,Witten:1998qj,Aharony:1999ti} and have a simple calculation. The entanglement entropy of a region $A$ in the boundary CFT is given by the area of a minimal surface $\mA$ in the bulk AdS space
\be
S_A=\f{\Area[\mA]}{4G},
\ee
with $G$ being the Newton constant. This is the Ryu-Takayanagi formula of holographic entanglement entropy \cite{Ryu:2006bv,Ryu:2006ef,Nishioka:2009un,Takayanagi:2012kg}. This is a classical gravity result, and one can also consider the quantum corrections \cite{Headrick:2010zt,Barrella:2013wja,Faulkner:2013ana}.

Quantum gravity in AdS$_3$ spacetime with cosmological constant $\L=-1/\ell^2$ is dual to a two-dimensional CFT with central charge \cite{Brown:1986nw}
\be
c=\f{3\ell}{2G}.
\ee
The small Newton constant expansion in gravity side corresponds to large central charge expansion in CFT side \cite{Headrick:2010zt,Hartman:2013mia,Faulkner:2013yia,Barrella:2013wja}.
The part of the R\'enyi entropy that is proportional to central charge is called classical, the next-to-leading part is called one-loop, and the next-to-next-to-leading part is called two-loop, and et.\ al.

The calculation of $N$-interval R\'enyi entropy in a two-dimensional CFT is equivalent to the calculation of a $2N$-point correlation function of twist operators \cite{Calabrese:2004eu}. For one interval on complex plane the R\'enyi entropy is universal \cite{Holzhey:1994we,Calabrese:2004eu}, but when for cases of two or more intervals there are no general results and the details of the CFT are relevant \cite{Caraglio:2008pk,Furukawa:2008,Calabrese:2009ez,Headrick:2010zt,Calabrese:2010he}. For two short intervals on a complex plane, on which we focus in this paper, one can calculate the R\'enyi entropy as expansion of the cross ratio $x$ in both gravity and CFT sides
\cite{Headrick:2010zt,Calabrese:2010he,Barrella:2013wja,Chen:2013kpa,Chen:2013dxa,Perlmutter:2013paa,Chen:2014kja,Beccaria:2014lqa,Headrick:2015gba,Zhang:2015hoa,Beccaria:2015shq}.
In CFT side one can use the OPE (operator product expansion) of twist operators, and various cases have been considered \cite{Headrick:2010zt,Calabrese:2010he,Chen:2013kpa,Chen:2013dxa,Perlmutter:2013paa,Chen:2014kja,Beccaria:2014lqa,Headrick:2015gba,Zhang:2015hoa}.
Using this method it is very cumbersome to calculate the R\'enyi entropy to higher order of the cross ratio $x$.
In gravity side the one-loop R\'enyi entropy of the graviton has been calculated to order $x^8$ in \cite{Barrella:2013wja}, and this result is reproduced in CFT side by considering contributions of stress tensor in \cite{Chen:2013kpa,Chen:2013dxa}.
There is a similar story for the one-interval R\'enyi entropy on a torus with the temperature being low or high \cite{Barrella:2013wja,Cardy:2014jwa,Chen:2014unl,Chen:2014ehg,Chen:2014hta,Chen:2015kua,Chen:2015uia,Chen:2016uvu}, but we will not consider the case in this paper.

It was pointed out in \cite{Beccaria:2014lqa} that if one takes the $n\to1$ limit and only calculates the entanglement entropy the calculation would be much easier, both in gravity and in CFT sides. In gravity side, the one-loop entanglement entropy of the graviton has been calculated to order $x^{10}$, that of the spin-3 field to order $x^{14}$, and that of the spin-4 field to order $x^{18}$ \cite{Beccaria:2014lqa}.
In this paper we adopt this strategy and calculate the one-loop entanglement entropy in CFT side. For stress tensor we calculate the one-loop entanglement entropy to order $x^{10}$, for $W_3$ operator to order $x^{12}$, and for $W_4$ operator to order $x^{14}$.

The rest of the paper is arranged as follows.
In Section~\ref{s2} we review the method of calculating the one-loop two-interval entanglement entropy from OPE of twist operators, as well as the gravity results that we want to reproduce in the CFT side.
In Section~\ref{s3} we calculate the contributions of stress tensor to the one-loop two-interval entanglement entropy.
In Section~\ref{s4} and~\ref{s5} we consider the cases $W_3$ operator and $W_4$ operator, respectively.
We end with conclusion and discussion in Section~\ref{s6}.
In Appendix~\ref{classoo} there are details of some general calculations that are useful to Section~\ref{s2}, \ref{s3}, and \ref{s4}.
In Appendix~\ref{sum} there are some summation formulas.

\section{Entanglement entropy from OPE of twist operators}\label{s2}

In this section we review small cross ratio expansion of entanglement entropy of two short intervals. We also give the basic setup of the calculation in the paper. It will be very brief here, and one may see details in \cite{Calabrese:2010he,Headrick:2010zt,Chen:2013kpa,Chen:2013dxa,Perlmutter:2013paa,Chen:2014kja,Beccaria:2014lqa}.

We consider a two-dimensional CFT on the complex plane, and the constant time slice is an infinite straight line. One can choose a subset $A$ of the line which is the union of several intervals, and name its complement as $B$. The vacuum state density matrix of the CFT is $\r=|0\rag\lag0|$, and one can trace out the degrees of freedom of $B$ and get the reduced density matrix
\be
\r_A=\tr_B\r.
\ee
For any positive integer $n>1$, one can define the order $n$ R\'enyi entropy
\be
S_A^{(n)}=-\f{1}{n-1}\log\tr_A\r_A^n.
\ee
For two subsets $A$ and $B$ that do not necessarily complement each other, one may define the R\'enyi mutual information
\be
I_{A,B}^{(n)} = S_A^{(n)} + S_B^{(n)} -S_{A\cup B}^{(n)}.
\ee
Taking the $n\to1$ limit one gets the entanglement entropy and mutual information.
\be
S_A=\lim_{n\to1}S_A^{(n)}, ~~~ I_{A,B}=\lim_{n\to1}I_{A,B}^{(n)}.
\ee

To get the order $n$ R\'enyi entropy of $N$ intervals, one uses the replica trick and calculates the partition function of the CFT on a genus $(n-1)(N-1)$ Riemann surface. This equals to the correlation function of $2N$ twist operators $\s$, $\td\s$ that are inserted at the boundaries of each interval on a complex plane in $\CFT^n$ that is the $n$-fold of the original CFT \cite{Calabrese:2004eu}. The twist operators $\s$, $\td\s$ are primary operators with conformal weights \cite{Calabrese:2004eu}
\be
h_\s=\bar h_\s=h_{\td\s}=\bar h_{\td\s}=\f{c(n^2-1)}{24n}.
\ee
For the case of two short intervals in a CFT where all the relevant operators can be decoupled as holomorphic and anti-holomorphic sectors and every anti-holomorphic operator is in one-to-one correspondence with a holomorphic one, one has the R\'enyi mutual information as a function of the cross ratio $x$ \cite{Calabrese:2010he,Headrick:2010zt,Chen:2013kpa,Chen:2013dxa,Perlmutter:2013paa,Chen:2014kja}
\be \label{mutualn}
I_n=\f{2}{n-1}\log\Big[ \sum_K \f{d_K^2}{\a_K} x^{h_K}{}_2F_1(h_K,h_K;2h_K;x) \Big].
\ee
Here the summation $K$ is over all the holomorphic linearly independent orthogonal quasiprimary operators $\Phi_K$ in $\CFT^n$, and every $\Phi_K$ is constructed from quasiprimary operators of the original CFT.
We call the quasiprimary operators in the original CFT as the old ones, and the quasiprimary operators in CFT$^n$ as the new ones.
Factor $\a_K$ is the normalization factor of $\Phi_K$
\be
\lag \Phi_K(z)\Phi_L(w) \rag_C=\f{\a_K\d_{KL}}{(z-w)^{2h_K}},
\ee
with $C$ denoting the complex plane. Factor $d_K$ is the OPE coefficient and it can be calculated as \cite{Calabrese:2010he}
\be \label{dK}
d_K=\f{1}{l^{h_K}}\lim_{z\to\inf}z^{2h_K}\lag\Phi_K(z)\rag_{\mR_{n,1}},
\ee
and here $\mR_{n,1}$ is an $n$-sheeted Riemann surface with the branch cut being $[0,l]$. The expectation value on $\mR_{n,1}$ with coordinate $z$ is calculated by mapping it to a complex plane with coordinate $f$ by
\be \label{fz}
f(z)=\Big( \f{z-l}{z} \Big)^{\f{1}{n}}.
\ee

When some new quasiprimary operators $\Phi_{K_p}$ with $p=1,2,\cdots,m$ in $\CFT^n$ are not orthogonal to each other, we can either orthogonalize them using Gram-Schmidt process, or just replace the summation of these operators to a product of two vectors and a matrix
\be
\f{d_K^2}{\a_K} \to d_K^\T \a_K^{-1} d_K.
\ee
Here $d_K^\T$ is the transpose of the $m$-dimensional vector $d_K$
\be
d_K^\T=(d_{K_1},d_{K_2},\cdots,d_{K_m}),
\ee
and $\a_K$ is the $m\times m$ normalization matrix
\be
\lag \Phi_{K_p} \Phi_{K_q} \rag_C=\f{\a_{K_{pq}}}{(z-w)^{2h_K}}, ~~~ p,q=1,2,\cdots,m,
\ee
and $\a_K^{-1}$ is the inverse of $\a_K$.

To calculate the R\'enyi mutual information (\ref{mutualn}) to higher order of $x$, one has to consider a large number of new quasiprimary operators, and this makes the method very cumbersome. However, it was shown in \cite{Beccaria:2014lqa} that if one is only interested in the mutual information, i.e.\ the $n\to1$ limit of the R\'enyi mutual information (\ref{mutualn}), the calculation can be simplified significantly. The example of contributions of scalar operators has been given therein. In this paper we will give more examples, including contributions of stress tensor, $W_3$ operator and $W_4$ operator. The mutual information is calculated as
\be \label{mutual}
I=\lim_{n\to1}\f{2}{n-1}\Big[ \sum_K \f{\hat d_K^2}{\a_K} x^{h_K}{}_2F_1(h_K,h_K;2h_K;x) \Big],
\ee
with $K$ denoting summation over the nonidentity holomorphic new quasiprimary operators of $\CFT^n$. Here $\hat d_K$ is got from $d_K$ by setting all the $n$'s, but the ones in trigonometric functions, to 1. It will not affect the result of mutual information, and it will make the calculation much simpler.
We will see in the subsequent sections of this paper that only some of new quasiprimary operators contribute to the mutual information.
Furthermore, the central charge $c$ dependence comes from ${\hat d_K^2}/{\a_K}$, and the number of new quasiprimary operators would be smaller if we only want to get the one-loop part of the mutual information.

The method of calculating the one-loop entanglement entropy in the gravity side was developed in \cite{Barrella:2013wja}. One uses the result in \cite{Maloney:2007ud,Yin:2007gv,Giombi:2008vd}, and calculate the 1-loop partition function in the background of the handlebody.\footnote{This gravity result has been recently justified in \cite{Chen:2015uga}.}
It has been calculated in gravity side that, the spin-2, spin-3, and spin-4 fields contribute to one-loop holographic mutual information
\bea \label{mutual234}
&& \hspace{-5mm} I_{\spin 2}^\oloop = \frac{x^4}{630}+\frac{2 x^5}{693}+\frac{15 x^6}{4004}+\frac{x^7}{234}+\frac{167 x^8}{36036}+\frac{69422 x^9}{14549535}+\frac{122 x^{10}}{24871}+O(x^{11}),          \nn\\
&& \hspace{-5mm} I_{\spin 3}^\oloop = \frac{x^6}{12012}+\frac{x^7}{4290}+\frac{7 x^8}{16830}+\frac{28 x^9}{46189}+\frac{15 x^{10}}{19019}+\frac{2 x^{11}}{2093}+\frac{1644627 x^{12}}{1487285800}+O(x^{13}),  \\
&& \hspace{-5mm} I_{\spin 4}^\oloop = \frac{x^8}{218790}+\frac{4 x^9}{230945}+\frac{3 x^{10}}{76076}+\frac{5 x^{11}}{71162}+\frac{11 x^{12}}{101660}+\frac{11 x^{13}}{72675}+\frac{1001 x^{14}}{5058180}+O(x^{15}).  \nn
\eea
In AdS/CFT correspondence, it is standard that the graviton corresponds to stress tensor in CFT side. Also a general spin-$s$ field in gravity side corresponds to $W_s$ and $\bar W_s$ operators in CFT side \cite{Henneaux:2010xg,Campoleoni:2010zq}.
In this paper we will reproduce the results (\ref{mutual234}) in the CFT side.

\section{Stress tensor}\label{s3}

In this section we consider an ordinary large central charge CFT, and we get the contributions of vacuum conformal family operators to the one-loop mutual information of two short intervals. The primary operator of the vacuum conformal family is the identity, and the holomorphic decedents are constructed by the stress tensor $T$, normal ordering and derivatives. Firstly we need to construct the new quasiprimary operators $\Phi_K$ we need, then we calculate the coefficients $\a_K$ and $\hat d_K$, and lastly we sum the results to get the mutual information.

\subsection{Construction of quasiprimary operators}

For the original CFT, we count the number of vacuum conformal family holomorphic operators in each level as
\be \label{chi2}
\chi_{(2)}=\tr_{(2)}x^{L_0} = \prod_{m=2}^\inf \f{1}{1-x^m}= 1+x^2+x^3+2 x^4+2 x^5+4 x^6+4 x^7+7 x^8+8 x^9+12 x^{10}+O(x^{11}).
\ee
Then the number of old holomorphic quasiprimary operators in each level is listed as
\be \label{e1}
(1-x)\chi_{(2)}+x=1+x^2+x^4+2 x^6+3 x^8+x^9+4 x^{10}+O(x^{11}).
\ee
They are listed in Table~\ref{quasiprimary2}.
At level 0, it is the identity operator 1. At level 2 we have the stress tensor $T$ and $\a_T=\f{c}{2}$. At level 4 we have
\be \mc A=(TT)-\f{3}{10}\p^2T, ~~~ \a_{\mc A}=\f{c(5c+22)}{10}. \ee
At level 6 we have
\bea
&& \mc B=(\p T\p T)-\f{4}{5}(T\p^2T)+\f{23}{210}\p^4T,                    \nn\\
&& \mc D=(T(TT))-\f{9}{10}(T\p^2 T)+\f{4}{35}\p^4 T+\f{93}{70c+29} \mc B, \\
&& \a_{\mc B}=\frac{36c (70 c+29)}{175}, ~~~
   \a_\mD=\frac{3 c (2 c-1) (5 c+22) (7 c+68)}{4 (70 c+29)}.              \nn
\eea
The quasiprimary operator $\mD$ is chosen such that the structure constant $C_{TT\mD}=0$, and $\mB$ is chosen such that it is orthogonal to $\mD$. At level 8 we have $\mA^{(8,m)}$ with $m=1,2,3$, and we need neither their explicit forms or their normalization factors. At level 9 we have $\mA^{(9)}$. At level 10 we have $\mA^{(10,m)}$ with $m=1,2,3,4$.

\begin{table}
  \centering
  \begin{tabular}{|c|c|c|c|c|c|c|c|c|}
     \hline
     level        & 0 & 2   & 4     & 6            & 8             & 9             & 10             & $\cdots$ \\\hline
     quasiprimary & 1 & $T$ & $\mA$ & $\mB$, $\mD$ & $\mA^{(8,m)}$ & $\mA^{(9)}$   & $\mA^{(10,m)}$ & $\cdots$ \\
     \hline
   \end{tabular}
  \caption{Old holomorphic quasiprimary operators of vacuum conformal family in the original CFT. The ranges in which the $m$'s take values can be seen easily in (\ref{e1}). At level 8 we have $m=1,2,3$, and at level 10 we have $m=1,2,3,4$.}\label{quasiprimary2}
\end{table}

Using the old holomorphic quasiprimary operators of the original CFT listed above as well as derivatives, we can construct all the new holomorphic quasiprimary operators of $\CFT^n$ to level 10.
Given $p$ old holomorphic quasiprimary operators of original CFT that are located at different replica $\mO_{j_1}$, $\mP_{j_2}$, $\mQ_{j_3}$, $\cdots$, we can just multiply them and get one new quasiprimary operator of $\CFT^n$
\be
\mO_{j_1}\mP_{j_2}\mQ_{j_3}\cdots.
\ee
Given also $q$ derivatives, we can get $C_{p+q-1}^q$ linearly independent operators, and so the number of linearly independent quasiprimary operators that can be constructed is
\be
C_{p+q-1}^q-C_{p+q-2}^{q-1}=C_{p+q-2}^q.
\ee
We denote these quasiprimary operators with one derivative as
\be
\I_m(\mO_{j_1}\mP_{j_2}\mQ_{j_3}\cdots), ~~~ m=1,2,\cdots,p-1,
\ee
or simply
\be
\I(\mO_{j_1}\mP_{j_2}\mQ_{j_3}\cdots).
\ee
For all the linearly independent new holomorphic quasiprimary operators with permutations of these $j_i$'s from 0 to $n-1$, we just denote them by
\be
\I(\mO\mP\mQ\cdots).
\ee
We use similar notations for the new holomorphic quasiprimary operators of $\CFT^n$ with two and more derivatives, and for example we have $\II(\mO\mP\mQ\cdots)$, $\III(\mO\mP\mQ\cdots)$, $\cdots$. We call these operators belong to the class $\mO\mP\mQ\cdots$.

The new holomorphic operators of $\CFT^n$ can be counted as $\chi_{(2)}^n$ with $\chi_{(2)}$ being defined in (\ref{chi2}), and so the new holomorphic quasiprimary operators can be counted as
\bea \label{counting2}
&& \hspace{-6mm}
   (1-x)\chi_{(2)}^n+x=1+n x^2+\frac{n (n+1)}{2} x^4+\frac{n(n-1)}{2} x^5+\frac{n (n+1) (n+5)}{6} x^6+\frac{n (n-1) (2 n+5)}{6} x^7 \nn\\
&& \hspace{-6mm}\phantom{(1-x)\chi_{(2)}^n+x=}
   +\frac{n (n+1) (n^2+17n+18)}{24} x^8+\frac{n (n+1) (3 n^2+19n-10)}{24} x^9    \\
&& \hspace{-6mm}\phantom{(1-x)\chi_{(2)}^n+x=}
   +\frac{n (n+1) (n^3+39n^2+156n+44)}{120} x^{10}+O(x^{11}).     \nn
\eea
We listed all these quasiprimary operators in Table~\ref{tabqua2}.

\begin{table}
  \centering
\begin{tabular}{|c|c|c|c|c|c|c|c|c|c|c|}\cline{1-5}\cline{7-11}
  level & quasiprimary  & ?? & \# & \#                                                                                                                && level             & quasiprimary           & ??                   & \#              & \#  \\ \cline{1-5}\cline{7-11}
  0 & 1 & \checked\checked & 1 & 1                                                                                                                    &&                   & $\I(\mA\mA)$           & \checked\checked     & $\f{n_2}{2}$ & \multirow{6}*{\rotatebox[origin=cc]{90}{$\f{n (n+1) (3 n^2+19n-10)}{24}$}} \\ \cline{8-10} \cline{1-5}
  2 & $T$ & \texttimes\texttimes & $n$ & $n$                                                                                                          &&                   & $\I(TT\mA)$            & \checked\checked     & $n_3$           &     \\ \cline{8-10} \cline{1-5}
  \multirow{2}*{4} & $\mA$ & \texttimes\texttimes & $n$            & \multirow{2}*{\rotatebox[origin=cc]{90}{$\f{n(n+1)}{2}$}}                          &&  \multirow{2}*{9} & $\I(TTTT)$             & \checked\checked     & $\f{n_4}{8}$    &     \\ \cline{8-10} \cline{2-4}
                   & $TT$  & \checked\checked     & $\f{n_2}{2}$ &                                                                                    &&                   & $\III(T\mA)$           & \texttimes\texttimes & $n_2$        & \\ \cline{8-10} \cline{1-5}
  5 & $\I(TT)$ & \checked\checked & $\f{n_2}{2}$  & $\f{n_2}{2}$                                                                                      &&                   & $\III(TTT)$            & \checked\texttimes   & $\f{2n_3}{3}$   & \\ \cline{8-10} \cline{1-5}
                   & $\mB$, $\mD$ & \texttimes\texttimes & $2n$  & \multirow{4}*{\rotatebox[origin=cc]{90}{$\frac{n (n+1) (n+5)}{6}$}}                &&                   & $\V(TT)$               & \checked\checked     & $\f{n_2}{2}$ & \\ \cline{7-11} \cline{2-4}
  \multirow{2}*{6} & $T\mA$       & \texttimes\texttimes & $n_2$ &                                                                                    &&     & $\mA^{(10,m)}$           & \texttimes\texttimes & $4n$            & \multirow{15}*{\rotatebox[origin=cc]{90}{$\f{n (n+1)(n^3+39n^2+156n+44)}{120}$}} \\ \cline{8-10} \cline{2-4}
                   & $TTT$        & \checked\texttimes   & $\f{n_3}{6}$ &                                                                             &&     & $T\mA^{(8,m)}$           & \texttimes\texttimes & $3n_2$       & \\ \cline{8-10} \cline{2-4}
                   & $\II(TT)$    & \checked\checked     & $\f{n_2}{2}$ &                                                                             &&     & $\mA\mB$, $\mA\mD$       & \texttimes\texttimes & $2n_2$       & \\ \cline{8-10} \cline{1-5}
    & $\I(T\mA)$ & \texttimes\texttimes & $n_2$        & \multirow{3}*{\rotatebox[origin=cc]{90}{$\frac{n (n-1) (2n+5)}{6}$}}                         &&     & $T\mA\mA$                & \checked\texttimes   & $\f{n_3}{2}$    & \\ \cline{8-10} \cline{2-4}
  7 & $\I(TTT)$  & \checked\texttimes   & $\f{n_3}{3}$ &                                                                                              &&     & $TT\mB$                  & \checked\checked     & $\f{n_3}{2}$    & \\ \cline{8-10} \cline{2-4}
    & $\III(TT)$ & \checked\checked     & $\f{n_2}{2}$ &                                                                                              &&     & $TT\mD$                  & \texttimes\texttimes & $\f{n_3}{2}$    & \\ \cline{8-10} \cline{1-5}
                   & $\mA^{(8,m)}$  & \texttimes\texttimes & $3n$          & \multirow{8}*{\rotatebox[origin=cc]{90}{$\f{n (n+1) (n^2+17n+18)}{24}$}} &&     & $TTT\mA$                 & \checked\texttimes   & $\f{n_4}{6}$    & \\ \cline{8-10} \cline{2-4}
                   & $T\mB$, $T\mD$ & \texttimes\texttimes & $2n_2$        &                                                                          &&  10 & $TTTTT$                  & \checked\texttimes   & $\f{n_5}{120}$  & \\ \cline{8-10} \cline{2-4}
                   & $\mA\mA$       & \checked\checked     & $\f{n_2}{2}$  &                                                                          &&     & $\II(T\mB)$, $\II(T\mD)$ & \texttimes\texttimes & $2n_2$       & \\ \cline{8-10} \cline{2-4}
  \multirow{2}*{8} & $TT\mA$        & \checked\checked     & $\f{n_3}{2}$  &                                                                          &&     & $\II(\mA\mA)$            & \checked\checked     & $\f{n_2}{2}$ & \\ \cline{8-10} \cline{2-4}
                   & $TTTT$         & \checked\checked     & $\f{n_4}{24}$ &                                                                          &&     & $\II(TT\mA)$             & \checked\checked     & $\f{3n_3}{2}$   & \\ \cline{8-10} \cline{2-4}
                   & $\II(T\mA)$    & \texttimes\texttimes & $n_2$         &                                                                          &&     & $\II(TTTT)$              & \checked\checked     & $\f{n_4}{4}$    & \\ \cline{8-10} \cline{2-4}
                   & $\II(TTT)$     & \checked\texttimes   & $\f{n_3}{2}$  &                                                                          &&     & $\IV(T\mA)$              & \texttimes\texttimes & $n_2$        &   \\ \cline{8-10} \cline{2-4}
                   & $\IV(TT)$      & \checked\checked     & $\f{n_2}{2}$  &                                                                          &&     & $\IV(TTT)$               & \checked\texttimes   & $\f{5n_3}{6}$   &   \\ \cline{8-10} \cline{1-5}
  \multirow{2}*{9} & $\mA^{(9)}$            & \texttimes\texttimes & $n$    &                                                                         &&     & $\VI(TT)$                & \checked\checked     & $\f{n_2}{2}$ &   \\ \cline{7-11} \cline{2-4}
                   & $\I(T\mB)$, $\I(T\mD)$ & \texttimes\texttimes & $2n_2$ &                                                                         && $\cdots$ & $\cdots$ & $\cdots$ & $\cdots$ & $\cdots$ \\ \cline{2-4} \cline{7-11}
\end{tabular}
\caption{All new holomorphic quasiprimary operators in CFT$^n$ to level 10. If we want to calculate the general order $n$ R\'enyi mutual information of two short intervals, we have to consider all of them. In the third column we marked the answers to two questions for the operators. The first question is whether the operators contribute to the mutual information, i.e. the order 1 R\'enyi mutual information, and the second question is whether it contribute to the one-loop part of the mutual information. If one answer is yes, we mark \checked, and if one answer is no, we mark \texttimes. In this paper we concentrate on the one-loop mutual information, and so we only need to consider the operators marked with two \checked's. In the fourth and fifth columns we count the degeneracies, and we have shorthand $n_m=n(n-1)\cdots(n-m+1)$. The counting is in accord with (\ref{counting2}).}
\label{tabqua2}
\end{table}

\subsection{Calculation of coefficients}

If we want to get the general R\'enyi mutual information using (\ref{mutualn}), we have to get coefficients $\a_K$ and $d_K$ for all the operators in Table~\ref{tabqua2}. In spirit of \cite{Beccaria:2014lqa}, after we take $n\to1$ limit and get the mutual information, only some of them contribute. A general old holomorphic quasiprimary operator $\mO$ with conformal weight $h$ transforms in an arbitrary conformal transformation $z\to f(z)$ as
\be \label{conftrans}
\mO(z)=f'^h\mO(f)+\cdots,
\ee
with $\cdots$ denoting terms that have the Schwarzian derivative or its derivatives. For the transformation (\ref{fz}) that we use to calculate $d_K$, the Schwarzian  derivative is proportional to $n-1$. We divide the nonidentity new quasiprimary operators of $\CFT^n$ in two cases.
\begin{itemize}
  \item For a new operator with only one nonidentity old quasiprimary operator of the original CFT in one replica, say $\mO_j$ with $j=0,1,\cdots,n-1$, coefficient $d_K$ only comes from the $\cdots$ in (\ref{conftrans}), and we have $d_K\sim n-1$. So the term $d_K^2/(n-1)$ vanishes in the $n \to 1$ limit. Such operators do not contribute to the mutual information.
  \item For the other cases, the coefficients $d_K$ is consisted by some trigonometric functions, and terms from $\cdots$ in (\ref{conftrans}) are still proportional to $n-1$. A summation of $d_K^2/\a_K$ is just a summation of some trigonometric functions, and this always leads to an overall factor $n-1$. After summation the contributions from $\cdots$ in (\ref{conftrans}) are proportional to $(n-1)^2$, and these terms over $n-1$ would vanish in the $n\to1$ limit.
\end{itemize}
From the above analysis, we need not the full form of $d_K$ to get the mutual information, we only need to replace $d_K$ by
\be \label{e30}
\hat d_K = d_K \textrm{~by~taking~all}~ n \to 1~\textrm{except~the~ones~in~trigonometric~functions}.
\ee
The new coefficient $\hat d_K$ is calculated using (\ref{dK}), (\ref{fz}), (\ref{e30}), as well as (\ref{conftrans}) without the $\cdots$.

To make the analysis of the large central charge limit easier, we define the modified normalization factor and the modified OPE coefficient
\be
\b_K = \lim_{c\to\inf} \f{\a_K}{\td\a_K}, ~~~ \hat b_K = \lim_{c\to\inf} \f{\hat d_K}{C_K},
\ee
with $\b_K$ and $\hat b_K$ being independent of the central charge. So we have
\be
\lim_{c\to\inf} \f{\hat d_K^2}{\a_K} = \bigg(\lim_{c\to\inf} \f{C_K^2}{\td\a_K} \bigg) \f{\hat b_K^2}{\b_K}.
\ee
For $\CFT^n$ quasiprimary operators with only one quasiprimary operator of the original CFT, we need not to consider them, as we have said above. For quasiprimary operators with two quasiprimary operators of the original CFT, we only need to consider the cases when the two operators are the same. We have the $\CFT^n$ operators of class $\mO\mO$
\be
\mO\mO, ~ \I(\mO\mO), ~ \II(\mO\mO), ~ \cdots.
\ee
For these operators we choose $C_K=\a_\mO$ and $\td\a_K=\a_\mO^2$. For quasiprimary operators with three quasiprimary operators of the original CFT, say class $\mO\mP\mQ$
\be
\mO\mP\mQ, ~ \I(\mO\mP\mQ), ~ \II(\mO\mP\mQ), ~ \cdots.
\ee
we choose $C_K=C_{\mO\mP\mQ}$ and $\td\a_K=\a_{\mO\mP\mQ}=\a_\mO\a_\mP\a_\mQ$ with $C_{\mO\mP\mQ}$ being the structure constant. For quasiprimary operators with four and more quasiprimary operators of the original CFT, usually we cannot make $\hat b_K$ and $\b_K$ independent of the central charge, but we can always make them independent of the central charge in the large central charge limit. Coefficients $C_K$ for these cases will be defined case by case. For all the quasiprimary operators in class $\mO\mP\mQ\cdots$, we have the coefficient $\td\a_K=\a_{\mO\mP\mQ\cdots}=\a_\mO\a_\mP\a_\mQ\cdots$.
With all these setups, we can easily identify whether some operators contribute to the mutual information or not, and if yes whether they contribute to the one-loop mutual information or not. The answers to the two questions are marked in the third column of Table~\ref{tabqua2}. The result is that we only need the quasiprimary operators of the classes $TT$, $\mA\mA$, $TT\mA$, $TTTT$, $TT\mB$ to get the one-loop mutual information.

For the classes of $TT$ and $\mA\mA$, the contributions to mutual information are just
\be
I_{TT}=I_{\mO\mO}|_{h=2}, ~~~ I_{\mA\mA}=I_{\mO\mO}|_{h=4},
\ee
with $I_{\mO\mO}$ being (\ref{IOO}).

For operators in class of $TT\mA$ we have the structure constant
\be
C_{TT\mA}=\f{c(5c+22)}{10}.
\ee
To level 10, the quasiprimary operators we need to consider are
\bea
&& TT\mA, ~~~ \I_1(TT\mA)=\ii\p TT\mA - T\ii\p T\mA, ~~~ \I_2(TT\mA)=\ii\p TT\mA -\f12 TT\ii\p \mA,  \nn\\
&& \II_1(TT\mA)=\p T\p T\mA - \f25 \p^2 TT\mA -\f25 T \p^2 T\mA,  \nn\\
&& \II_2(TT\mA)=\p T T\p\mA - \f45 \p^2 TT\mA -\f29 T T\p^2\mA,\\
&& \II_3(TT\mA)= T\p T\p\mA - \f45 T\p^2 T\mA -\f29 T T\p^2\mA.\nn
\eea
We have the modified normalization factors
\be
\b_{TT\mA}=1, ~~~
\b_{\sI(TT\mA)}= 2 \lt(\begin{array}{cc} 4 & 2 \\ 2 & 3 \end{array}\rt), ~~~
\b_{\sII(TT\mA)}=\f{16}{45} \lt(\ba{ccc} 81 & 36 & 36 \\ 36 & 182 & 20 \\ 36 & 20 & 182 \ea\rt).
\ee
The OPE coefficients are
\bea
&& \hat b_{TT\mA}^{j_1j_2j_3}=\f{1}{2^8}\f{1}{s_{j_1j_3}^4 s_{j_2j_3}^4},  ~~~
   \hat b_{\sI_1(TT\mA)}^{j_1j_2j_3}=\f{1}{2^7}\f{c_{j_1j_2}}{s_{j_1j_3}^5 s_{j_2j_3}^5},       \nn\\
&& \hat b_{\sI_2(TT\mA)}^{j_1j_2j_3}=\f{1}{2^8}\f{s_{j_1j_2}-2s_{j_1j_2j_3}}{s_{j_1j_3}^5 s_{j_2j_3}^5},   ~~~
   \hat b_{\sII_1(TT\mA)}^{j_1j_2j_3}=-\f{1}{2^7}\f{s_{j_1j_2}^2}{s_{j_1j_3}^6 s_{j_2j_3}^6},   \nn\\
&& \hat b_{\sII_2(TT\mA)}^{j_1j_2j_3}=\f{1}{9 \cdot 2^8}\f{ 10 s_{j_1j_2}^2 -36s_{j_1j_3}^2 +45s_{j_2j_3}^2 + 36 s_{j_1j_2j_3}^2}
                                                          {s_{j_1j_3}^6 s_{j_2j_3}^6},     \\
&& \hat b_{\sII_3(TT\mA)}^{j_1j_2j_3}=\f{1}{9 \cdot 2^8}\f{ 10 s_{j_1j_2}^2 +45s_{j_1j_3}^2 -36s_{j_2j_3}^2 + 36 s_{j_1j_2j_3}^2}
                                                          {s_{j_1j_3}^6 s_{j_2j_3}^6},     \nn
\eea
with the definitions $s_{j_1j_2}=\sin(\f{j_1-j_2}{n}\pi)$, $s_{j_1j_2j_3}=\sin(\f{j_1+j_2-2j_3}{n}\pi)$, $c_{j_1j_2}=\cos(\f{j_1-j_2}{n}\pi)$ and the ones similar to them.

For operators in class $TTTT$, we choose
\be
C_K=\f{c^2}{4}.
\ee
To level 10, we need the operators
\bea
&& TTTT, ~~~
   \I_1(TTTT)=\ii\p TTTT - T\ii\p TTT,  \nn\\
&& \I_2(TTTT)=\ii\p TTTT - TT\ii\p TT, ~~~
   \I_3(TTTT)=\ii\p TTTT - TTT\ii\p T,  \nn\\
&& \II_1(TTTT)=\p T\p T TT -\f25\p^2TTTT-\f25T\p^2 TTT, \nn\\
&& \II_2(TTTT)=\p TT\p TT -\f25\p^2TTTT-\f25T T\p^2 TT, \nn\\
&& \II_3(TTTT)=\p TT T\p T -\f25\p^2TTTT-\f25T TT\p^2 T, \\
&& \II_4(TTTT)=T\p T\p TT -\f25T\p^2TTT-\f25T T\p^2 TT, \nn\\
&& \II_5(TTTT)=T\p T T\p T -\f25T\p^2TTT-\f25T TT\p^2 T, \nn\\
&& \II_6(TTTT)=TT\p T\p T -\f25TT\p^2TT-\f25T TT\p^2 T. \nn
\eea
The modified normalization factors are
\be
\b_{TTTT}=1,
~~~
\b_{\sI(TTTT)}=4
\left(
\begin{array}{ccc}
 2 & 1 & 1 \\
 1 & 2 & 1 \\
 1 & 1 & 2
\end{array}
\right),
~~~
\b_{\sII(TTTT)}=\f{16}{5}
\left(
\begin{array}{cccccc}
 9 & 2 & 2 & 2 & 2 & 0 \\
 2 & 9 & 2 & 2 & 0 & 2 \\
 2 & 2 & 9 & 0 & 2 & 2 \\
 2 & 2 & 0 & 9 & 2 & 2 \\
 2 & 0 & 2 & 2 & 9 & 2 \\
 0 & 2 & 2 & 2 & 2 & 9
\end{array}
\right).
\ee
We need the leading part of the four-point function
\be
\lag T(f_1)T(f_2)T(f_3)T(f_4)\rag_C=\f{c^2}{4}\Big(   \f{1}{f_{12}^4f_{34}^4}
                                                    + \f{1}{f_{13}^4f_{24}^4}
                                                    + \f{1}{f_{14}^4f_{23}^4}  \Big)+O(c),
\ee
with the definition $f_{12}=f_1-f_2$ and the ones similar to it.
The modified OPE coefficients are
\bea
&& \hat b_{TTTT}^{j_1j_2j_3j_4}=\f{1}{2^8} \Big(  \f{1}{s_{j_1j_2}^4s_{j_3j_4}^4}
                                                + \f{1}{s_{j_1j_3}^4s_{j_2j_4}^4}
                                                + \f{1}{s_{j_1j_4}^4s_{j_2j_3}^4} \Big),                         \nn\\
&& \hat b_{\sI_1(TTTT)}^{j_1j_2j_3j_4}=\f{1}{2^7} \Big(   \f{s_{j_1j_3j_2j_4}}{s_{j_1j_3}^5 s_{j_2j_4}^5}
                                                       + \f{s_{j_1j_4j_2j_3}}{s_{j_1j_4}^5 s_{j_2j_3}^5}
                                                       - \f{2c_{j_1j_2}}{s_{j_1j_2}^5 s_{j_3j_4}^4} \Big),       \nn\\
&& \hat b_{\sI_2(TTTT)}^{j_1j_2j_3j_4} = \hat b_{\sI_1(TTTT)}^{j_1j_3j_2j_4}, ~~~
   \hat b_{\sI_3(TTTT)}^{j_1j_2j_3j_4} = \hat b_{\sI_1(TTTT)}^{j_1j_4j_2j_3},                                    \\
&& \hat b_{\sII_1(TTTT)}^{j_1j_2j_3j_4} = \f{1}{2^7} \Big(   \f{s_{j_1j_3j_2j_4}^2}{s_{j_1j_3}^6s_{j_2j_4}^6}
                                                           + \f{s_{j_1j_4j_2j_3}^2}{s_{j_1j_4}^6s_{j_2j_3}^6}
                                                           + \f{9-8s_{j_1j_2}^2}{2s_{j_1j_2}^6s_{j_3j_4}^4} \Big), \nn\\
&& \hat b_{\sII_2(TTTT)}^{j_1j_2j_3j_4} = \hat b_{\sII_1(TTTT)}^{j_1j_3j_2j_4}, ~~~
   \hat b_{\sII_3(TTTT)}^{j_1j_2j_3j_4} = \hat b_{\sII_1(TTTT)}^{j_1j_4j_2j_3},  \nn\\
&& \hat b_{\sII_4(TTTT)}^{j_1j_2j_3j_4} = \hat b_{\sII_1(TTTT)}^{j_2j_3j_1j_4}, ~~~
   \hat b_{\sII_5(TTTT)}^{j_1j_2j_3j_4} = \hat b_{\sII_1(TTTT)}^{j_2j_4j_1j_3}, ~~~
   \hat b_{\sII_6(TTTT)}^{j_1j_2j_3j_4} = \hat b_{\sII_1(TTTT)}^{j_3j_4j_1j_2}.  \nn
\eea
Here there are new definition $s_{j_1j_3j_2j_4}=\sin(\f{j_1-j_3-j_2+j_4}{n}\pi)$ and the ones similar to it.

For operators in class $TT\mB$, we have the structure constant
\be
C_{TT\mB}=-\f{2c(70c+29)}{35}, ~~~
\ee
and the operators, modified normalization factors, and modified OPE coefficients are
\be
TT\mB, ~~~ \b_{TT\mB}=1, ~~~ \hat b_{TT\mB}^{j_1j_2j_3}=-\f{1}{2^{10}}\f{s_{j_1j_2}^2}{s_{j_1j_3}^6 s_{j_2j_3}^6}.
\ee

\subsection{One-loop mutual information}

Using the coefficients in the last subsection and the summation formulas in Appendix~\ref{sum} we can get the one-loop mutual information. The contributions from operators of class $TT$, $\mA\mA$, $TT\mA$, $TTTT$, and $TT\mB$ are respectively
\bea
&& \hspace{-5mm}
   I_{TT}^{\oloop}=\frac{x^4}{630}+\frac{2 x^5}{693}+\frac{15 x^6}{4004}+\frac{x^7}{234}+\frac{7 x^8}{1530}
                   +\frac{84 x^9}{17765}+\frac{x^{10}}{209}+O(x^{11}),                                               \nn\\
&& \hspace{-5mm}
   I_{\mA\mA}^{\oloop}=\frac{x^8}{218790}+\frac{4 x^9}{230945}+\frac{3 x^{10}}{76076}+O(x^{11}),                      ~~~
   I_{TT\mA}^{\oloop}= -\frac{x^8}{109395}-\frac{8 x^9}{230945}-\frac{3 x^{10}}{38038}+O(x^{11}),                     \nn\\
&& \hspace{-5mm}
   I_{TTTT}^{\oloop}= \frac{x^8}{15708}+\frac{878 x^9}{14549535}+\frac{207 x^{10}}{1293292}+O(x^{11}),                ~~~
   I_{TT\mB}^{\oloop}= O(x^{11}).
\eea
Summing them together, we get the contributions of the vacuum conformal family to one-loop mutual information
\be
I_{(2)}^\oloop = \frac{x^4}{630}+\frac{2 x^5}{693}+\frac{15 x^6}{4004}+\frac{x^7}{234}+\frac{167 x^8}{36036}+\frac{69422 x^9}{14549535}+\frac{122 x^{10}}{24871}+O(x^{11}),
\ee
and this matches the gravity result in \cite{Beccaria:2014lqa}, i.e. $I_{\spin 2}^\oloop$ in (\ref{mutual234}). Note that $I_{TT}^{\oloop}$ matches $I_{(2)}^\oloop$ to order $x^7$.

\section{$W_3$ operator}\label{s4}

In a CFT with $W(2,3)$ symmetry, there are operators $W$ with conformal weights (3,0) and $\bar W$ with conformal weights (0,3) besides the operators $T$ and $\bar T$. In such a CFT  the contributions from the stress tensor still exist. In this section we consider the additional contributions to the one-loop mutual information because of the existence of the $W$ operator.

\subsection{Construction of quasiprimary operators}

We count the holographic operators in the original CFT with $W(2,3)$ symmetry as
\be \label{chi23}
\chi_{(2,3)}=\tr_{(2,3)} x^{L_0}=\prod_{m=0}^\inf\f{1}{1-x^{m+2}}\f{1}{1-x^{m+3}}.
\ee
The holomorphic quasiprimary operators are counted as
\be
(1-x)\chi_{(2,3)}+x= 1+x^2+x^3+x^4+x^5+4 x^6+2 x^7+7 x^8+7 x^9+12 x^{10}+14 x^{11}+26 x^{12}+O(x^{13}),
\ee
and the additional ones compared to an ordinary CFT are counted as
\be
(1-x)(\chi_{(2,3)}-\chi_{(2)})= x^3+x^5+2 x^6+2 x^7+4 x^8+6 x^9+8 x^{10}+12 x^{11}+19 x^{12}+O(x^{13}),
\ee
with $\chi_{(2)}$ being defined in (\ref{chi2}). The holomorphic operators in the conformal family of a general holomorphic nonidentity primary operator $\phi$ with conformal weights $(h,0)$ are counted as
\be \label{chiphi}
\chi_\phi=\tr_\phi x^{L_0}=x^h \chi, ~~~ \chi \equiv \prod_{m=1}^\inf\f{1}{1-x^{m}}.
\ee

The number of quasiprimary operators in conformal family of $\phi$ is counted as
\be \label{e2}
(1-x)\chi_\phi=x^h \left[1+x^2+x^3+2 x^4+2 x^5+4 x^6+4 x^7+7 x^8+8 x^9+12 x^{10}+O(x^{11})\right].
\ee
When $\phi$ is the operator $W$ we have $h=3$, and we choose $\a_W=\f{c}{3}$ as usual. At level 5, we have the quasiprimary operator
\be
\mU=(TW)-\f{3}{14}\p^2W, ~~~ \a_\mU=\frac{c(7 c+114)}{42}.
\ee
At level 6, we have
\be
\mV=(T\ii\p W)-\f32 (\ii\p T W)-\f18 \ii\p^3W, ~~~ \a_\mV=\frac{5 c (c+2)}{2}.
\ee
At level 7, we have two quasiprimary operators
\bea
&& \mX=(\p T\p W)-\f{2}{7}(T\p^2W)-\f35 (\p^2T W)+\f{1}{42}\p^4W,                        \nn\\
&& \mZ=(T(TW))-\f37(T\p^2W)-\f3{10} (\p^2T W)+\f{1}{28}\p^4W + \frac{141}{35 c+53} \mX,  \\
&& \a_\mX=\frac{264 c (35 c+53)}{1225}, ~~~
   \a_\mZ=\frac{c (c+23) (5 c-4) (7 c+114)}{6 (35 c+53)}.                                \nn
\eea
Here $\mZ$ is chosen such that the structure constant $C_{TW\mZ}=0$, and $\mX$ is chosen such that it is orthogonal to $\mZ$. We also have the useful structure constants
\bea
C_{TW\mU}=\frac{c (7 c+114)}{42}, ~~~
C_{TW\mV}=-\ii c (c+2), ~~~
C_{TW\mX}=-\frac{2c (35 c+53)}{35}.
\eea

The additional holomorphic primary operators in the original CFT with $W(2,3)$ symmetry are counted as
\be
\f{\chi_{(2,3)}-\chi_{(2)}}{\chi}=x^3+x^6+x^8+x^9+x^{10}+x^{11}+3 x^{12}+O(x^{13}),
\ee
with $\chi_{(2)}$ in (\ref{chi2}), $\chi_{(2,3)}$ in (\ref{chi23}), and $\chi$ in (\ref{chiphi}).
At level 3, it is just $W$, and at level 6, 8, 9, 10, 11 we name them $\mE$, $\mF$, $\mG$, $\mH$, and $\mI$, respectively. At level 12, there are three of them, and we name them $\mJ$, $\mK$, $\mL$. We list them and their decedent quasiprimary operators in Table~\ref{quasi23tab}. The explicit form of $\mE$ can be found in, for example, the review \cite{Bouwknegt:1992wg}, from which we can get
\be
\a_\mE=\f{2c^2}{9}+O(c), ~~~ C_{WW\mE} = \f{2c^2}{9}+O(c).
\ee
The explicit forms, normalization factors, structure constants of other primary operators will not be used in this paper.

\begin{table}
  \centering
  \begin{tabular}{|c|c|c|c|c|c|c|c|c|c|c|c|c|}
   \hline
   $L_0$               & 2   & 3   & 4     & 5     & 6             & 7            & 8             & 9           & 10             & 11             & 12                  & $\cdots$ \\ \hline
   \#                  & 1   & 1   & 1     & 1     & 4             & 2            & 7             & 7           & 12             & 14             & 26                  & $\cdots$ \\ \hline
   1                   & $T$ &     & $\mA$ &       & $\mA^{(6,m)}$ &              & $\mA^{(8,m)}$ & $\mA^{(9)}$ & $\mA^{(10,m)}$ & $\mA^{(11,m)}$ & $\mA^{(12,m)}$      & $\cdots$ \\ \hline
   $W$                 &     & $W$ &       & $\mU$ & $\mV$         & $\mX$, $\mZ$ & $W^{(8,m)}$   & $W^{(9,m)}$ & $W^{(10,m)}$   & $W^{(11,m)}$   & $W^{(12,m)}$        & $\cdots$ \\ \hline
   $\mE$               &     &     &       &       & $\mE$         &              & $\mE^{(8)}$   & $\mE^{(9)}$ & $\mE^{(10,m)}$ & $\mE^{(11,m)}$ & $\mE^{(12,m)}$      & $\cdots$ \\ \hline
   $\mF$               &     &     &       &       &               &              & $\mF$         &             & $\mF^{(10)}$   & $\mF^{(11)}$   & $\mF^{(12,m)}$      & $\cdots$ \\ \hline
   $\mG$               &     &     &       &       &               &              &               & $\mG$       &                & $\mG^{(11)}$   & $\mG^{(12)}$        & $\cdots$ \\ \hline
   $\mH$               &     &     &       &       &               &              &               &             & $\mH$          &                & $\mH^{(12)}$        & $\cdots$ \\ \hline
   $\mI$               &     &     &       &       &               &              &               &             &                & $\mI$          &                     & $\cdots$ \\ \hline
   $\mJ$, $\mK$, $\mL$ &     &     &       &       &               &              &               &             &                &                & $\mJ$, $\mK$, $\mL$ & $\cdots$ \\ \hline
  \end{tabular}
  \caption{Old holographic nonidentity quasiprimary operators in the original CFT with $W(2,3)$ symmetry. In the first line, there are the levels. In second line it is the number of quasiprimary operators in each level. From the third line, we list the quasiprimary operators in each conformal family, and the primary operator for each conformal family is given at the first column. There are some $m$'s in the table, and they take values in different ranges. The range that each $m$ takes values can be figured out in (\ref{e1}) and (\ref{e2}). }\label{quasi23tab}
\end{table}

The additional new holomorphic quasiprimary operators in $\CFT^n$ with $W(2,3)$ symmetry compared with an ordinary $\CFT^n$, are counted as
\bea \label{e32}
&& (1-x)(\chi_{(2,3)}^n-\chi_{(2)}^n)          =    n x^3+n^2 x^5+\frac{n (3 n+1)}{2} x^6+\frac{n (n^2+4n-1)}{2} x^7 +\frac{n (n+1) (3 n+1)}{2} x^8  \nn\\
&& \phantom{(1-x)(\chi_{(2,3)}^n-\chi_{(2)}^n) =}  +\frac{n (n+1) (n^2+18n-1)}{6} x^9 +\frac{n (9 n^3+58n^2+27n+2)}{12} x^{10}           \nn\\
&& \phantom{(1-x)(\chi_{(2,3)}^n-\chi_{(2)}^n) =}  +\frac{n (n^4+52n^3+179n^2+68n-12)}{24} x^{11}                                        \\
&& \phantom{(1-x)(\chi_{(2,3)}^n-\chi_{(2)}^n) =}  +\frac{n (6 n^4+109n^3+232n^2+83n+26)}{24} x^{12}+O(x^{13}).                         \nn
\eea
They are listed in Table~\ref{quasi23ntab}.

\newcommand{\rot}[2]{\multirow{#1}*{\rotatebox[origin=cc]{90}{#2}}}

\begin{table}
\centering
\begin{tabular}{|c|c|c|c|c|c|c|c|c|c|c|}\cline{1-5}\cline{7-11}
  $L_0$ & quasiprimary & ??? & \# & \#                                                                                                                                  &&      $L_0$ & quasiprimary & ??? & \# & \#                                                                                                                                                \\\cline{7-11}\cline{1-5}
  3 & $W$ & \texttimes\texttimes\texttimes & $n$ & $n$                                                                                                                  &&     & $TW^{(9,m)}$, $TE^{(9)}$,           &                                               &                              & \rot{13}{$\f{n(n^4+52n^3+179n^2+68n-12)}{24}$}                \\  \cline{1-5}
  \multirow{2}*{5} & $\mU$ & \texttimes\texttimes\texttimes & $n$   & \multirow{2}*{$n^2$}                                                                              &&     & $T\mG$, $\mA\mX$, $\mA\mZ$,         & \texttimes\texttimes\texttimes                & $13n_2$                      &                                                               \\  \cline{2-4}
                   & $TW$  & \texttimes\texttimes\texttimes & $n_2$ &                                                                                                   &&     & $\mA^{(6,m)}\mU$, $\mA^{(8,m)}W$    &                                               &                              &                                                               \\\cline{8-10}  \cline{1-5}
    & $\mV$, $\mE$ & \texttimes\texttimes\texttimes & $2n$          & \rot{3}{$\frac{n (3 n+1)}{2}$}                                                                    &&     & $WW^{(8,m)}$, $\mU\mV$              & \checked\texttimes\texttimes                  & $3n_2$                       &                                                               \\\cline{8-10}  \cline{2-4}
  6 & $WW$         & \checked\checked\checked       & $\f{n_2}{2}$ &                                                                                                    &&     & $W\mE^{(8)}$, $W\mF$, $\mU\mE$      & \texttimes\texttimes\texttimes                & $3n_2$                       &                                                               \\\cline{8-10}  \cline{2-4}
    & \multicolumn{3}{c|}{$n_2$}                                   &                                                                                                    &&     & $TT\mX$, $TT\mZ$, $T\mA\mU$,        & \multirow{2}*{\texttimes\texttimes\texttimes} & \multirow{2}*{$\f{9n_3}{2}$} &                                                               \\  \cline{1-5}
                   & $\mX$, $\mZ$    & \texttimes\texttimes\texttimes & $2n$         & \rot{4}{$\f{n (n^2+4n-1)}{2}$}                                                   &&  11 & $T\mA^{(6,m)}W$, $\mA\mA W$         &                                               &                              &                                                               \\\cline{8-10}  \cline{2-4}
  \multirow{2}*{7} & $T\mU$, $\mA W$ & \texttimes\texttimes\texttimes & $2n_2$       &                                                                                  &&     & $TW\mV$                             & \checked\checked\checked                      & $n_3$                        &                                                               \\\cline{8-10}  \cline{2-4}
                   & $TTW$           & \texttimes\texttimes\texttimes & $\f{n_3}{2}$ &                                                                                  &&     & $TW\mE$, $WW\mU$                    & \texttimes\texttimes\texttimes                & $\f{n_3}{2}$                 &                                                               \\\cline{8-10}  \cline{2-4}
                   & \multicolumn{3}{c|}{$\frac{3n_2}{2}$}                           &                                                                                  &&     & $TTT\mU$, $TT\mA W$                 & \texttimes\texttimes\texttimes                & $\f{2n_4}{3}$                &                                                               \\\cline{8-10}  \cline{1-5}
    & $W^{(8,m)}$, $\mE^{(8)}$, $\mF$ & \texttimes\texttimes\texttimes & $4n$         & \rot{5}{$\f{n (n+1) (3 n+1)}{2}$}                                               &&     & $TWWW$                              & \texttimes\texttimes\texttimes                & $\f{n_4}{6}$                 &                                                               \\\cline{8-10}  \cline{2-4}
    & $T\mV$, $T\mE$                  & \texttimes\texttimes\texttimes & $2n_2$       &                                                                                 &&     & $TTTTW$                             & \texttimes\texttimes\texttimes                & $\f{n_5}{24}$                &                                                               \\\cline{8-10}  \cline{2-4}
  8 & $W\mU$                          & \checked\texttimes\texttimes   & $n_2$        &                                                                                 &&     & \multicolumn{3}{c|}{$\frac{n(n-1)(9 n^2+67 n+94)}{12}$}                                                            &                                                               \\\cline{7-11}  \cline{2-4}
    & $TWW$                           & \checked\checked\texttimes     & $\f{n_3}{2}$ &                                                                                 &&     & $W^{(12,m)}$, $\mE^{(12,m)}$,                  &                                               &                              & \rot{23}{$\f{n (6 n^4+109n^3+232n^2+83n+26)}{24}$} \\  \cline{2-4}
    & \multicolumn{3}{c|}{$\frac{n(n-1)(2 n+3)}{2}$}                                  &                                                                                 &&     & $\mF^{(12,m)}$, $\mG^{(12)}$,                  & \texttimes\texttimes\texttimes                & $19n$                        &                                                    \\  \cline{1-5}
  \multirow{8}*{9} & $W^{(9,m)}$, $\mE^{(9)}$, $\mG$ & \texttimes\texttimes\texttimes                & $6n$                  & \rot{8}{$\frac{n (n+1) (n^2+18n-1)}{6}$} &&     & $\mH^{(12)}$, $\mJ$, $\mK$, $\mL$              &                                               &                              &                                                    \\\cline{8-10}  \cline{2-4}
                   & $T\mX$, $T\mZ$, $\mA\mU$,       & \multirow{2}*{\texttimes\texttimes\texttimes} & \multirow{2}*{$5n_2$} &                                          &&     & $TW^{(10,m)}$, $TE^{(10,m)}$,                  & \multirow{4}*{\texttimes\texttimes\texttimes} & \multirow{4}*{$17n_2$}       &                                                    \\
                   & $\mA^{(6,m)}W$                  &                                               &                       &                                          &&     & $T\mF^{(10)}$, $T\mH$, $\mA W^{(8,m)}$,        &                                               &                              &                                                    \\  \cline{2-4}
                   & $W\mV$                          & \checked\texttimes\texttimes                  & $n_2$                 &                                          &&     & $\mA \mE^{(8)}$, $\mA\mF$, $\mA^{(6,m)}\mV$,   &                                               &                              &                                                    \\  \cline{2-4}
                   & $W\mE$                          & \texttimes\texttimes\texttimes                & $n_2$                 &                                          &&     & $\mA^{(6,m)}\mE$, $\mA^{(9)}W$                 &                                               &                              &                                                    \\\cline{8-10}  \cline{2-4}
                   & $TT\mU$, $T\mA W$, $WWW$        & \texttimes\texttimes\texttimes                & $\f{5n_3}{3}$         &                                          &&     & $WW^{(9,m)}$, $\mU\mX$, $\mU\mZ$               & \checked\texttimes\texttimes                  & $6n_2$                       &                                                    \\\cline{8-10}  \cline{2-4}
                   & $TTTW$                          & \texttimes\texttimes\texttimes                & $\f{n_4}{6}$          &                                          &&     & $W\mE^{(9)}$, $W\mG$, $\mV\mE$                 & \texttimes\texttimes\texttimes                & $3n_2$                       &                                                    \\\cline{8-10}  \cline{2-4}
                   & \multicolumn{3}{c|}{$\frac{n(n-1)(5n+3)}{2}$}                                                           &                                          &&     & $\mV\mV$, $\mE\mE$                             & \checked\checked\checked                      & $n_2$                        &                                                    \\\cline{8-10}  \cline{1-5}
     & $W^{(10,m)}$, $\mE^{(10,m)}$, & \multirow{2}*{\texttimes\texttimes\texttimes} & \multirow{2}*{$8n$}   & \rot{11}{$\frac{n (9 n^3+58n^2+27n+2)}{12}$}             &&     & $TTW^{(8,m)}$, $TT\mE^{(8)}$,                  & \multirow{2}*{\texttimes\texttimes\texttimes} & \multirow{2}*{$4n_3$}        &                                                    \\
     & $\mF^{(10)}$, $\mH$           &                                               &                       &                                                          &&  12 & $TT\mF$, $T\mA\mV$, $T\mA\mE$                  &                                               &                              &                                                    \\\cline{8-10}  \cline{2-4}
     & $TW^{(8,m)}$, $T\mE^{(8)}$,   & \multirow{2}*{\texttimes\texttimes\texttimes} & \multirow{2}*{$6n_2$} &                                                          &&     & $\mA^{(6,m)}WW$,                             & \multirow{2}*{\checked\checked\texttimes}      & \multirow{2}*{$\f{5n_3}{2}$} &                                                     \\
     & $T\mF$, $\mA\mV$, $\mA\mE$    &                                               &                       &                                                          &&     & $\mA W\mU$, $T\mU\mU$                          &                                               &                              &                                                    \\\cline{8-10}  \cline{2-4}
     & $W\mX$, $W\mZ$                & \checked\texttimes\texttimes                  & $2n_2$                &                                                          &&     & $TW\mX$                                        & \checked\checked\checked                      & $n_3$                        &                                                    \\\cline{8-10}  \cline{2-4}
  10 & $\mU\mU$                      & \checked\checked\checked                      & $\f{n_2}{2}$          &                                                          &&     & $TW\mZ$, $WW\mV$                               & \texttimes\texttimes\texttimes                & $\f{3n_3}{2}$                        &                                            \\\cline{8-10}  \cline{2-4}
     & $TT\mV$, $TT\mE$              & \texttimes\texttimes\texttimes                & $n_3$                 &                                                          &&     & $WW\mE$                                        & \checked\checked\checked                      & $\f{n_3}{2}$                 &                                                    \\\cline{8-10}  \cline{2-4}
     & $TW\mU$                       & \checked\checked\checked                      & $n_3$                 &                                                          &&     & $TTT\mV$, $TTT\mE$                             & \texttimes\texttimes\texttimes                & $\f{n_4}{3}$                 &                                                    \\\cline{8-10}  \cline{2-4}
     & $\mA WW$                      & \checked\checked\texttimes                    & $\f{n_3}{2}$          &                                                          &&     & $TTW\mU$, $T\mA WW$                            & \checked\checked\texttimes                    & $n_4$                        &                                                    \\\cline{8-10}  \cline{2-4}
     & $TTWW$                        & \checked\checked\checked                      & $\f{n_4}{4}$          &                                                          &&     & $WWWW$                                         & \checked\checked\checked                      & $\f{n_4}{24}$                &                                                    \\\cline{8-10}  \cline{2-4}
     & \multicolumn{3}{c|}{$\frac{n(n-1)(3n^2+26n+17)}{6}$}                                                  &                                                          &&     & $TTTWW$                                        & \checked\checked\texttimes                    & $\f{n_5}{12}$                &                                                    \\\cline{8-10}  \cline{1-5}
  \multirow{2}*{11} & $W^{(11,m)}$, $\mE^{(11,m)}$,       & \multirow{2}*{\texttimes\texttimes\texttimes} & \multirow{2}*{$12n$}         &                              &&     & \multicolumn{3}{c|}{$\frac{n(n-1)(n^3+53n^2+232n+300)}{24}$}                                                                  &                                                    \\\cline{7-11}
                    & $\mF^{(11)}$, $\mG^{(11)}$, $\mI$   &                                               &                              &                              &&     $\cdots$ &  $\cdots$ & $\cdots$ & $\cdots$ & $\cdots$                                                                                                                                \\\cline{7-11}   \cline{2-4}
\end{tabular}
  \caption{Additional new holographic quasiprimary operators in $\CFT^n$ with $W(2,3)$ symmetry. The operators with derivatives can be constructed from the ones without derivatives easily, and so we only list the number of such operators in each level. In the third column we mark whether the operators contribute to the R\'enyi mutual information $I_n$, mutual information $I$, and one-loop part of mutual information $I_\oloop$. The counting in the fourth and fifth columns is in accord with (\ref{e32}).}\label{quasi23ntab}
\end{table}

\subsection{Calculation of coefficients}

To level 12, the new holomorphic quasiprimary operators in $\CFT^n$ that contribute to the one-loop mutual information are the ones in classes $WW$, $\mU\mU$, $TW\mU$, $TTWW$, $TW\mV$, $\mV\mV$, $\mE\mE$, $TW\mX$, $WW\mE$, and $WWWW$. The contributions of operators in classes $WW$, $\mU\mU$, $\mV\mV$, and $\mE\mE$ are
\be
I_{WW}=I_{\mO\mO}|_{h=3}, ~~~
I_{\mU\mU}=I_{\mO\mO}|_{h=5}, ~~~
I_{\mV\mV}=I_{\mE\mE}=I_{\mO\mO}|_{h=6},
\ee
with $I_{\mO\mO}$ in (\ref{IOO}).

For operators in class $TW\mU$, we have
\bea
&& TW\mU, ~~~ I_1(TW\mU)=\ii\p TW\mU-\f23 T\ii\p W\mU, ~~~ I_2(TW\mU)=\ii\p T W\mU-\f25 TW\ii\p \mU, \nn\\
&& \II_1(TW\mU)=\p T\p W\mU-\f35\p^2TW\mU-\f27T\p^2W\mU,   \\
&& \II_2(TW\mU)=\p TW\p \mU-\p^2TW\mU-\f2{11}TW\p^2\mU,   \nn\\
&& \II_3(TW\mU)=T\p W \p\mU-\f57T\p^2W\mU-\f3{11}TW\p^2\mU,   \nn
\eea
the modified normalization factors
\be
\b_{TW\mU} = 1, ~~~
\b_{\sI(TW\mU)} = \f{4}{15}\left(\begin{array}{cc} 25 & 15 \\ 15 & 21\end{array}\right), ~~~
\b_{\sII(TW\mU)} = \frac{12}{385} \left( \begin{array}{ccc}  1452 & 770 & 550 \\  770 & 2800 & 350 \\  550 & 350 & 3825 \end{array} \right),
\ee
and the modified OPE coefficients
\bea
&& \hat b_{TW\mU}^{j_1j_2j_3}=-\f{1}{2^{10}}\f{1}{s_{j_1j_3}^4 s_{j_2j_3}^6},                                                    ~~~
   \hat b_{\sI_1(TW\mU)}^{j_1j_2j_3} = -\f{1}{2^{9}} \f{s_{j_1j_2}}{s_{j_1j_3}^5 s_{j_2j_3}^7},                                  \nn\\
&& \hat b_{\sI_2(TW\mU)}^{j_1j_2j_3} = -\f{1}{5\cdot2^{9}} \f{2s_{j_1j_2}-5s_{j_1j_2j_3}}{s_{j_1j_3}^5 s_{j_2j_3}^7},             ~~~
   \hat b_{\sII_1(TW\mU)}^{j_1j_2j_3} = -\f{3}{2^{10}} \f{s_{j_1j_2}^2}{s_{j_1j_3}^6 s_{j_2j_3}^8},                               \nn\\
&& \hat b_{\sII_2(TW\mU)}^{j_1j_2j_3} = -\f{1}{11\cdot2^{10}} \f{10s_{j_1j_2}^2-44s_{j_1j_3}^2+55s_{j_2j_3}^2+55s_{j_1j_2j_3}^2}
                                                                {s_{j_1j_3}^6 s_{j_2j_3}^8},                                      \\
&& \hat b_{\sII_3(TW\mU)}^{j_1j_2j_3} = -\f{3}{11\cdot2^{11}} \f{21s_{j_1j_2}^2+77s_{j_1j_3}^2-66s_{j_2j_3}^2+55s_{j_1j_2j_3}^2}
                                                                {s_{j_1j_3}^6 s_{j_2j_3}^8}.                                      \nn
\eea
In class $TW\mV$, we have operators
\be
TW\mV, ~~~ \I_1(TW\mV)=\ii\p T W\mV-\f23 T\ii\p W\mV, ~~~ \I_2(TW\mV)=\ii\p T W\mV-\f13 TW\ii\p \mV,
\ee
the modified normalization factors
\be
\b_{TW\mV}=1, ~~~ \b_{\sI(TW\mV)}=\frac{4}{3} \left(\begin{array}{cc} 5 & 3 \\ 3 & 4 \end{array}\right),
\ee
and the modified OPE coefficients
\bea
&& \hat b_{TW\mV}^{j_1j_2j_3}=\f{\ii}{2^{11}}\f{s_{j_1j_2}}{s_{j_1j_3}^5 s_{j_2j_3}^7}, ~~~
   \hat b_{\sI_1(TW\mV)}^{j_1j_2j_3} = \f{\ii}{3\cdot2^{12}} \f{12s_{j_1j_2}^2 +2s_{j_1j_3}^2 +3s_{j_2j_3}^2}
                                                               {s_{j_1j_3}^6 s_{j_2j_3}^8},                      \nn\\
&& \hat b_{\sI_2(TW\mV)}^{j_1j_2j_3} = \f{\ii}{3\cdot2^{12}} \f{5s_{j_1j_2}^2 -12s_{j_1j_3}^2 +15s_{j_2j_3}^2}
                                                               {s_{j_1j_3}^6 s_{j_2j_3}^8}.
\eea
For operators in class $TW\mX$, we have
\be
TW\mX, ~~~ \b_{TW\mX}=1, ~~~ \hat b_{TW\mX}^{j_1j_2j_3}=\f{1}{2^{12}}\f{s_{j_1j_2}^2}{s_{j_1j_3}^6 s_{j_2j_3}^8}.
\ee
For operators in class $WW\mE$, we have
\be
WW\mE, ~~~ \b_{WW\mE}=1, ~~~ \hat b_{WW\mE}^{j_1j_2j_3}=\f{1}{2^{12}}\f{1}{s_{j_1j_3}^6 s_{j_2j_3}^6}.
\ee

In class $TTWW$, we choose $C_K=\f{c^2}{6}$ and we have
\bea
&& TTWW, ~~~ \I_1(TTWW)=\ii\p TTWW-T\ii\p TWW,                                            \nn\\
&& \I_2(TTWW)=\ii\p TTWW- \f23 TT\ii\p WW, ~~~ \I_3(TTWW)=\ii\p TTWW - \f23 TTW\ii\p W,    \nn\\
&& \II_1(TTWW)=\p T \p TWW - \f25 \p^2 TTWW - \f25 T \p^2TWW,                             \nn\\
&& \II_2(TTWW)=\p T T\p WW - \f35 \p^2 TTWW - \f27 TT\p^2WW,                             \nn\\
&& \II_3(TTWW)=\p T TW\p W - \f35 \p^2 TTWW - \f27 TTW\p^2W,                             \\
&& \II_4(TTWW)=T \p T\p WW - \f35 T\p^2 TWW - \f27 TT\p^2WW,                             \nn\\
&& \II_5(TTWW)= T \p TW\p W - \f35 T\p^2 TWW - \f27 TTW\p^2W,                             \nn\\
&& \II_6(TTWW)=T T\p W\p W - \f37 TT\p^2 WW - \f37 TTW\p^2W,                             \nn
\eea
modified normalization factors
\be
\b_{TTWW}=1, ~~~
\b_{\sI(TTWW)}=\frac{4}{3} \left(\begin{array}{ccc} 6 & 3 & 3 \\ 3 & 5 & 3 \\ 3 & 3 & 5\end{array}\right), ~~~
\b_{\sII(TTWW)}=\frac{12}{35}
\left(\begin{array}{cccccc}
 84 & 28 & 28 & 28 & 28 & 0 \\
 28 & 132 & 42 & 20 & 0 & 30 \\
 28 & 42 & 132 & 0 & 20 & 30 \\
 28 & 20 & 0 & 132 & 42 & 30 \\
 28 & 0 & 20 & 42 & 132 & 30 \\
 0 & 30 & 30 & 30 & 30 & 195
\end{array}\right),
\ee
and modified OPE coefficients
\bea
&& \hat b_{TTWW}^{j_1j_2j_3j_4}=-\f{1}{2^{10}}\f{1}{s_{j_1j_2}^4s_{j_3j_4}^6},                           ~~~
   \hat b_{\sI_1(TTWW)}^{j_1j_2j_3j_4}=\f{1}{2^{8}}\f{c_{j_1j_2}}{s_{j_1j_2}^5s_{j_3j_4}^6},             \nn\\
&& \hat b_{\sI_2(TTWW)}^{j_1j_2j_3j_4}=-\f{1}{2^{9}}\f{s_{j_1j_2j_3j_4}}{s_{j_1j_2}^5s_{j_3j_4}^7},      ~~~
   \hat b_{\sI_3(TTWW)}^{j_1j_2j_3j_4}=\hat b_{\sI_2(TTWW)}^{j_1j_2j_4j_3},                              \nn\\
&& \hat b_{\sII_1(TTWW)}^{j_1j_2j_3j_4}=-\f{1}{2^{10}}\f{9-8s_{j_1j_2}^2}{s_{j_1j_2}^6s_{j_3j_4}^6},     ~~~
   \hat b_{\sII_2(TTWW)}^{j_1j_2j_3j_4}=-\f{3}{2^{10}}\f{s_{j_1j_2j_3j_4}^2}{s_{j_1j_2}^6s_{j_3j_4}^8},  \\
&& \hat b_{\sII_3(TTWW)}^{j_1j_2j_3j_4} = \hat b_{\sII_2(TTWW)}^{j_1j_2j_4j_3},                          ~~~
   \hat b_{\sII_4(TTWW)}^{j_1j_2j_3j_4} = \hat b_{\sII_2(TTWW)}^{j_2j_1j_3j_4},                          \nn\\
&& \hat b_{\sII_5(TTWW)}^{j_1j_2j_3j_4} = \hat b_{\sII_2(TTWW)}^{j_2j_1j_4j_3},                          ~~~
   \hat b_{\sII_6(TTWW)}^{j_1j_2j_3j_4}=-\f{3}{2^{11}}\f{13-12s_{j_3j_4}^2}{s_{j_1j_2}^4s_{j_3j_4}^8}.   \nn
\eea
For operators in class $WWWW$, we choose $C_K=\f{c^2}{9}$ and we have
\be
WWWW, ~~~
\b_{WWWW}=1, ~~~
\hat b_{WWWW}^{j_1j_2j_3j_4}=\f{1}{2^{12}}\Big( \f{1}{s_{j_1j_2}^6s_{j_3j_4}^6}
                                               +\f{1}{s_{j_1j_3}^6s_{j_2j_4}^6}
                                               +\f{1}{s_{j_1j_4}^6s_{j_2j_3}^6} \Big).
\ee

\subsection{One-loop mutual information}

Using the coefficients in the last subsection and the summation formulas in Appendix~\ref{sum} we can get the one-loop mutual information. The contributions from operators of different classes are respectively
\bea
&& \hspace{-5mm}
   I_{WW}^{\oloop} = \frac{x^6}{12012}+\frac{x^7}{4290}+\frac{7 x^8}{16830}
                    +\frac{28 x^9}{46189}+\frac{15 x^{10}}{19019}
                    +\frac{2 x^{11}}{2093}+\frac{33 x^{12}}{29900}+O(x^{13}),                                  \nn\\
&& \hspace{-5mm}
   I_{\mU\mU}^{\oloop} = \frac{x^{10}}{3879876}+\frac{5 x^{11}}{4056234}+\frac{33 x^{12}}{9657700}+O(x^{13}), ~~~
   I_{TW\mU}^\oloop = -\frac{x^{10}}{1939938}-\frac{5 x^{11}}{2028117}-\frac{33 x^{12}}{4828850}+O(x^{13}),     \nn\\
&& \hspace{-5mm}
   I_{TTWW}^\oloop = \frac{x^{10}}{3879876}+\frac{5 x^{11}}{4056234}+\frac{58 x^{12}}{16900975}+O(x^{13}),  ~~~
   I_{TW\mV}^\oloop = -\frac{x^{12}}{33801950}+O(x^{13}),                                                      \\
&& \hspace{-5mm}
   I_{\mV\mV}^\oloop = I_{\mE\mE}^\oloop = \frac{x^{12}}{67603900}+O(x^{13}), ~~~
   I_{TW\mX}^\oloop = O(x^{13}),                                                                               \nn\\
&& \hspace{-5mm}
   I_{WW\mE}^\oloop = -\frac{x^{12}}{33801950}+O(x^{13}), ~~~
   I_{WWWW}^\oloop = \frac{3163 x^{12}}{1487285800}+O(x^{13}).                                                 \nn
\eea
Summing them together, we get the additional contributions of the $W_3$ operator to one-loop mutual information
\be
I_{(3)}^\oloop = \frac{x^6}{12012}+\frac{x^7}{4290}+\frac{7 x^8}{16830}+\frac{28 x^9}{46189}+\frac{15 x^{10}}{19019}+\frac{2 x^{11}}{2093}+\frac{1644627 x^{12}}{1487285800}+O(x^{13}),
\ee
and this matches the gravity result in \cite{Beccaria:2014lqa}, i.e. $I_{\spin 3}^\oloop$ in (\ref{mutual234}). Note that $I_{WW}^{\oloop}$ matches $I_{(3)}^\oloop$ to order $x^{11}$. We also find that there is cancellation
\be \label{cancel1}
I_{\mU\mU}^{\oloop}+I_{TW\mU}^{\oloop}+I_{TTWW}^{\oloop}+I_{TW\mV}^{\oloop}+I_{\mV\mV}^{\oloop}+I_{TW\mX}^{\oloop}=O(x^{13}).
\ee

\section{$W_4$ operator}\label{s5}

The case of CFT with $W(2,4)$ symmetry is similar to the CFT with $W(2,3)$ symmetry.
In a CFT with $W(2,4)$ symmetry, there are operators $W$ with conformal weights (4,0) and $\bar W$ with conformal weights (0,4), besides the stress tensor $T$ and $\bar T$.

\subsection{Construction of quasiprimary operators}

The old holomorphic operators in the CFT with $W(2,4)$ symmetry are counted as
\be
\chi_{(2,4)}=\prod_{m=0}^\inf\f{1}{1-x^{m+2}}\f{1}{1-x^{m+4}},
\ee
among which the quasiprimary ones are counted as
\be
(1-x)\chi_{(2,4)}+x=1+x^2+2 x^4+3 x^6+x^7+6 x^8+3 x^9+10 x^{10}+7 x^{11}+19 x^{12}+14 x^{13}+32 x^{14}+O(x^{15}).
\ee
The nonidentity holomorphic primary operators are counted as
\be
\f{\chi_{(2,4)}-\chi_{(2)}}{\chi}=x^4+x^8+x^{10}+2 x^{12}+2 x^{14}+O(x^{15}),
\ee
with $\chi_{(2)}$ being defined in (\ref{chi2}) and $\chi$ being defined in (\ref{chiphi}). At level 4, it is just $W$, at level 8 we denote it by $\mE$, at level 10 we denote it by $\mF$, at level 12  we denote them by $\mG$ and $\mH$, and at level 14 we denote them by $\mI$ and $\mJ$.
As usual we choose $\a_W=\f{c}{4}$.
In conformal family of $W$, at level 6 we have the quasiprimary operator
\be
\mU=(TW)-\f16 \p^2W, ~~~ \a_\mU=\frac{c(c+24)}{8},
\ee
at level 7 we have the quasiprimary operator
\be
\mV=(T\ii\p W)-2 (\ii T W)-\f{1}{10}\ii\p^3W, ~~~ \a_\mV=\frac{3 c (5 c+22)}{5},
\ee
at level 9 we have the two quasiprimary operators
\bea
&& \hspace{-5mm}
   \mX = (\p T\p W)-\f29 (T\p^2 W) -\f45 (\p^2T W)+\f{1}{66}\p^4W,                      \nn\\
&& \hspace{-5mm}
   \mZ = (T(TW))-\f13(T\p^2W)-\f{3}{10}(\p^2TW)+\f{1}{44}\p^4W + \f{273}{55c+137} \mX,  \\
&& \hspace{-5mm}
   \a_\mX = \frac{364 c (55 c+137)}{2475}, ~~~
   \a_\mZ = \frac{c (c+24) (c+31) (55 c-6)}{8 (55 c+137)}.                              \nn
\eea
Here $\mZ$ is chosen such that the structure constant $C_{TW\mZ}=0$, and $\mX$ is chosen such that it is orthogonal to $\mZ$.
The structure constants that will be useful are
\be
C_{TW\mU}=\frac{c(c+24)}{8}, ~~~ C_{TW\mV}=-\frac{\ii c(5c+22)}{5}, ~~~ C_{TW\mX}=-\frac{2c(55c+137)}{55}.
\ee
To level 14 the old holomorphic quasiprimary operators are listed in Table~\ref{quasi24tab}.

\begin{table}
  \centering
  \begin{tabular}{|c|c|c|c|c|c|c|c|c|c|c|c|c|}
   \hline
   $L_0$        & 2   & 4     & 6             & 7     & 8             & 9           & 10             & 11              & 12              & 13             & 14             & $\cdots$ \\ \hline
   \#           & 1   & 2     & 3             & 1     & 6             & 3           & 10             & 7               & 19              & 14             & 32             & $\cdots$ \\ \hline
   1            & $T$ & $\mA$ & $\mA^{(6,m)}$ &       & $\mA^{(8,m)}$ & $\mA^{(9)}$ & $\mA^{(10,m)}$ & $\mA^{(11,m)}$  & $\mA^{(12,m)}$  & $\mA^{(13,m)}$ & $\mA^{(14,m)}$ & $\cdots$ \\ \hline
   $W$          &     & $W$   & $\mU$         & $\mV$ & $\mX$, $\mZ$  & $W^{(9,m)}$ & $W^{(10,m)}$   & $W^{(11,m)}$    & $W^{(12,m)}$    & $W^{(13,m)}$   & $W^{(14,m)}$   & $\cdots$ \\ \hline
   $\mE$        &     &       &               &       & $\mE$         &             & $\mE^{(10)}$   & $\mE^{(11)}$    & $\mE^{(12,m)}$  & $\mE^{(13,m)}$ & $\mE^{(14,m)}$ & $\cdots$ \\ \hline
   $\mF$        &     &       &               &       &               &             & $\mF$          &                 & $\mF^{(12)}$    & $\mF^{(13)}$   & $\mF^{(14,m)}$ & $\cdots$ \\ \hline
   $\mG$        &     &       &               &       &               &             &                &                 & $\mG$           &                & $\mG^{(14)}$   & $\cdots$ \\ \hline
   $\mH$        &     &       &               &       &               &             &                &                 & $\mH$           &                & $\mH^{(14)}$   & $\cdots$ \\ \hline
   $\mI$, $\mJ$ &     &       &               &       &               &             &                &                 &                 &                & $\mI$, $\mJ$   & $\cdots$ \\ \hline
  \end{tabular}
  \caption{Old holographic nonidentity quasiprimary operators in the original CFT with $W(2,4)$ symmetry.}\label{quasi24tab}
\end{table}

The additional new holomorphic quasiprimary operators in $\CFT^n$ are counted as
\bea
&& (1-x)(\chi_{(2,4)}^n-\chi_{(2)}^n) =
   n x^4+n^2 x^6+n^2 x^7 + \frac{n(n^2+4n+1)}{2} x^8 + \frac{n(2n^2+3n-1)}{2} x^9                  \nn\\
&& \phantom{(1-x)(\chi_{(2,4)}^n-\chi_{(2)}^n) =}
  + \frac{n(n+1)(n^2+14n+3)}{6} x^{10} + \frac{n (n^3+7n^2+3n-1)}{2} x^{11}                        \\
&& \phantom{(1-x)(\chi_{(2,4)}^n-\chi_{(2)}^n) =}
  + \frac{n (n^4+36n^3+147n^2+84n+20)}{24} x^{12} + \frac{n (2 n^4+35n^3+86n^2+19n-10)}{12} x^{13}  \nn\\
&& \phantom{(1-x)(\chi_{(2,4)}^n-\chi_{(2)}^n) =}
  + \frac{n (n^5+70n^4+695n^3+1310n^2+504n+60)}{120} x^{14}+O(x^{15}),                              \nn
\eea
and they are listed in Table~\ref{quasi24ntab}.

\begin{table}
  \centering
\begin{tabular}{|c|c|c|c|c|c|c|c|c|c|c|}\cline{1-5}\cline{7-11}
  $L_0$ & quasiprimary & ??? & \# & \#                                                                                                                                                 &&  $L_0$ & quasiprimary & ??? & \# & \#                                                                                                                                                                                \\ \cline{7-11} \cline{1-5}
  4 & $W$ & \texttimes\texttimes\texttimes & $n$ & $n$                                                                                                                                 &&                    & $TTT\mU$, $TT\mA W$                  & \texttimes\texttimes\texttimes                & $\f{2n_4}{3}$        &                                                                                                  \\ \cline{8-10} \cline{1-5}
  \multirow{2}*{6} & $\mU$ & \texttimes\texttimes\texttimes & $n$     & \multirow{2}*{$n^2$}                                                                                           &&  \multirow{2}*{12} & $TTWW$                               & \checked\checked\checked                      & $\f{n_4}{4}$         &                                                                                                  \\ \cline{8-10} \cline{2-4}
                   & $TW$  & \texttimes\texttimes\texttimes & $n_2$   &                                                                                                                &&                    & $TTTTW$                              & \texttimes\texttimes\texttimes                & $\f{n_5}{24}$        &                                                                                                  \\ \cline{8-10} \cline{1-5}
  \multirow{2}*{7} & $\mV$ & \texttimes\texttimes\texttimes & $n$ & \multirow{2}*{$n^2$}                                                                                               &&                    & \multicolumn{3}{c|}{$\frac{n(n-1)(2n^2+9n+5)}{2}$}                                                          &                                                                                                  \\ \cline{7-11} \cline{2-4}
                   & \multicolumn{3}{c|}{$n_2$}                   &                                                                                                                    &&                    & $W^{(13,m)}$, $\mE^{(13,m)}$, $\mF^{(13)}$        & \texttimes\texttimes\texttimes                & $11n$                  & \rot{8}{$\f{n(2n^4+35n^3+86n^2+19n-10)}{12}$}                      \\ \cline{8-10} \cline{1-5}
    & $\mX$, $\mZ$, $\mE$ & \texttimes\texttimes\texttimes & $3n$         & \rot{5}{$\frac{n(n^2+4n+1)}{2}$}                                                                           &&                    & $TW^{(11,m)}$, $T\mE^{(11)}$, $\mA W^{(9,m)}$,    & \multirow{2}*{\texttimes\texttimes\texttimes} & \multirow{2}*{$10n_2$} &                                                                    \\ \cline{2-4}
    & $T\mU$, $\mA W$     & \texttimes\texttimes\texttimes & $2n_2$       &                                                                                                            &&                    & $\mA^{(6,m)}\mV$, $\mA^{(9)}W$,                   &                                               &                        &                                                                    \\ \cline{8-10} \cline{2-4}
  8 & $WW$                & \checked\checked\checked       & $\f{n_2}{2}$ &                                                                                                            &&  \multirow{2}*{13} & $WW^{(9,m)}$, $\mU\mV$                            & \checked\texttimes\texttimes                  & $3n_2$                 &                                                                    \\ \cline{8-10} \cline{2-4}
    & $TTW$               & \texttimes\texttimes\texttimes & $\f{n_3}{2}$ &                                                                                                            &&                    & $TTW^{(9,m)}$, $T\mA\mV$                          & \texttimes\texttimes\texttimes                & $2n_3$                 &                                                                    \\ \cline{8-10} \cline{2-4}
    & \multicolumn{3}{c|}{$n_2$}                                          &                                                                                                            &&                    & $TW\mV$                                           & \checked\checked\checked                      & $n_3$                  &                                                                    \\ \cline{8-10} \cline{1-5}
    & $W^{(9,m)}$ & \texttimes\texttimes\texttimes & $2n$  & \rot{3}{$\f{n(2n^2+3n-1)}{2}$}                                                                                            &&                    & $TTT\mV$                                          & \texttimes\texttimes\texttimes                & $\f{n_4}{6}$           &                                                                    \\ \cline{8-10} \cline{2-4}
  9 & $T\mV$      & \texttimes\texttimes\texttimes & $n_2$ &                                                                                                                           &&                    & \multicolumn{3}{c|}{$\frac{n(n-1)(2n^3+35n^2+97n+46)}{12}$}                                                                &                                                                    \\ \cline{7-11} \cline{2-4}
    & \multicolumn{3}{c|}{$\frac{n(2n^2+n-3)}{2}$}         &                                                                                                                           &&                    & $W^{(14,m)}$, $\mE^{(14,m)}$, $\mF^{(14,m)}$,        & \multirow{2}*{\texttimes\texttimes\texttimes} & \multirow{2}*{$22n$}          & \rot{26}{$\f{n(n^5+70n^4+695n^3+1310n^2+504n+60)}{120}$} \\ \cline{1-5}
     & $W^{(10,m)}$, $\mE^{(10)}$, $\mF$                & \texttimes\texttimes\texttimes                 & $6n$                  & \rot{8}{$\f{n(n+1)(n^2+14n+3)}{6}$}                 &&                    & $\mG^{(14)}$, $\mH^{(15)}$, $\mI$, $\mJ$             &                                               &                               &                                                          \\ \cline{8-10} \cline{2-4}
     & $T\mX$, $T\mZ$, $T\mE$,                          & \multirow{2}*{\texttimes\texttimes\texttimes}  & \multirow{2}*{$6n_2$} &                                                     &&                    & $TW^{(12,m)}$, $T\mE^{(12,m)}$,                      & \multirow{5}*{\texttimes\texttimes\texttimes} & \multirow{5}*{$31n_2$}        &                                                          \\
     & $\mA\mU$, $\mA^{(6,m)}W$                         &                                                &                       &                                                     &&                    & $T\mF^{(12)}$, $T\mG$, $T\mH$, $\mA W^{(10,m)}$,     &                                               &                               &                                                          \\ \cline{2-4}
     & $W\mU$                                           & \checked\texttimes\texttimes                   & $n_2$                 &                                                     &&                    & $\mA\mE^{(10)}$, $\mA\mF$, $\mA^{(6,m)}\mX$,         &                                               &                               &                                                          \\ \cline{2-4}
  10 & $TT\mU$, $T\mA W$                                & \texttimes\texttimes\texttimes                 & $\f{3n_3}{2}$         &                                                     &&                    & $\mA^{(6,m)}\mZ$, $\mA^{(6,m)}\mE$,                  &                                               &                               &                                                          \\ \cline{2-4}
     & $TWW$                                            & \checked\checked\texttimes                     & $\f{n_3}{2}$          &                                                     &&                    & $\mA^{(8,m)}\mU$, $\mA^{(10,m)}W$                    &                                               &                               &                                                          \\ \cline{8-10} \cline{2-4}
     & $TTTW$                                           & \texttimes\texttimes\texttimes                 & $\f{n_4}{6}$          &                                                     &&                    & $WW^{(10,m)}$, $\mU\mX$, $\mU\mZ$                    & \checked\texttimes\texttimes                  & $6n_2$                        &                                                          \\ \cline{8-10} \cline{2-4}
     & \multicolumn{3}{c|}{$\frac{3n(n^2-1)}{2}$}                                                        &                                                                             &&                    & $W\mE^{(10)}$, $W\mF$, $\mU\mE$                      & \texttimes\texttimes\texttimes                & $3n_2$                        &                                                          \\ \cline{8-10} \cline{1-5}
     & $W^{(11,m)}$, $\mE^{(11)}$ & \texttimes\texttimes\texttimes & $5n$         & \rot{5}{$\frac{n (n^3+7n^2+3n-1)}{2}$}                                                             &&                    & $\mV\mV$                                             & \checked\checked\checked                      & $\f{n_2}{2}$                  &                                                          \\ \cline{8-10} \cline{2-4}
     & $TW^{(9,m)}$, $\mA\mV$     & \texttimes\texttimes\texttimes & $3n_2$       &                                                                                                    &&                    & $TTW^{(10,m)}$, $TT\mE^{(10)}$,                      & \multirow{4}*{\texttimes\texttimes\texttimes} & \multirow{4}*{$\f{27n_3}{2}$} &                                                          \\ \cline{2-4}
  11 & $W\mV$                     & \checked\texttimes\texttimes   & $n_2$        &                                                                                                    &&                    & $TT\mF$, $T\mA\mX$, $T\mA\mZ$, $T\mA\mE$,            &                                               &                               &                                                          \\ \cline{2-4}
     & $TTV$                      & \texttimes\texttimes\texttimes & $\f{n_3}{2}$ &                                                                                                    &&  \multirow{2}*{14} & $T\mA^{(6,m)}\mU$, $T\mA^{(8,m)}W$,                  &                                               &                               &                                                          \\ \cline{2-4}
     & \multicolumn{3}{c|}{$\frac{n(n-1)(n^2+7n+5)}{2}$}                          &                                                                                                    &&                    & $\mA\mA\mU$, $\mA\mA^{(6,m)}W$                       &                                               &                               &                                                          \\ \cline{8-10} \cline{1-5}
     & $W^{(12,m)}$, $\mE^{(12,m)}$         & \multirow{2}*{\texttimes\texttimes\texttimes} & \multirow{2}*{$12n$}  & \rot{13}{$\f{n (n^4+36n^3+147n^2+84n+20)}{24}$}                  &&                    & $TW\mX$                                              & \checked\checked\checked                      & $n_3$                         &                                                          \\ \cline{8-10}
     & $\mF^{(12)}$, $\mG$, $\mH$           &                                               &                       &                                                                  &&                    & $TW\mZ$, $TW\mE$                                     & \texttimes\texttimes\texttimes                & $2n_3$                        &                                                          \\ \cline{8-10} \cline{2-4}
     & $TW^{(10,m)}$, $T\mE^{(10)}$, $T\mF$ &                                               &                       &                                                                  &&                    & $T\mU\mU$, $\mA W\mU$, $\mA^{(6,m)}WW$               & \checked\checked\texttimes                    & $\f{5n_3}{2}$                 &                                                          \\ \cline{8-10}
     & $\mA\mX$, $\mA\mZ$, $\mA\mE$,        & \texttimes\texttimes\texttimes                & $14n_2$               &                                                                  &&                    & $WW\mU$                                              & \checked\checked\texttimes                    & $\f{n_3}{2}$                  &                                                          \\ \cline{8-10}
     & $\mA^{(6,m)}\mU$, $\mA^{(8,m)}W$     &                                               &                       &                                                                  &&                    & $TTT\mX$, $TTT\mZ$, $TTT\mE$                         &                                               &                               &                                                          \\ \cline{2-4}
     & $W\mX$, $W\mZ$                       & \checked\texttimes\texttimes                  & $2n_2$                &                                                                  &&                    & $TT\mA\mU$, $TT\mA^{(6,m)}W$,                        & \texttimes\texttimes\texttimes                & $\f{5n_4}{2}$                 &                                                          \\ \cline{2-4}
  12 & $W\mE$                               & \texttimes\texttimes\texttimes                & $n_2$                 &                                                                  &&                    & $T\mA\mA W$, $TTW\mU$,                               &                                               &                               &                                                          \\ \cline{2-4} \cline{8-10}
     & $\mU\mU$                             & \checked\checked\checked                      & $\f{n_2}{2}$          &                                                                  &&                    & $T\mA WW$, $TWWW$                                    & \checked\checked\texttimes                    & $\f{7n_4}{6}$                 &                                                          \\ \cline{8-10} \cline{2-4}
     & $TT\mX$, $TT\mZ$,                    &                                               &                       &                                                                  &&                    & $TTTT\mU$, $TTT\mA W$                                & \texttimes\texttimes\texttimes                & $\f{5n_5}{24}$                &                                                          \\ \cline{8-10}
     & $TT\mE$, $T\mA\mU$,                  & \texttimes\texttimes\texttimes                &  $5n_3$               &                                                                  &&                    & $TTTWW$                                              & \checked\checked\texttimes                    & $\f{n_5}{12}$                 &                                                          \\ \cline{8-10}
     & $T\mA^{(6,m)}$W, $\mA\mA W$          &                                               &                       &                                                                  &&                    & $TTTTTW$                                             & \texttimes\texttimes\texttimes                & $\f{n_6}{120}$                &                                                          \\ \cline{8-10} \cline{2-4}
     & $TW\mU$                              & \checked\checked\checked                      & $n_3$                 &                                                                  &&                    & \multicolumn{3}{c|}{$\f{n(n-1)(n+2)(5n^2+47n+24)}{12}$}                                                                              &                                                          \\ \cline{7-11} \cline{2-4}
     & $\mA WW$, $WWW$                      & \checked\checked\texttimes                    & $\f{2n_3}{3}$         &                                                                  &&  $\cdots$ & $\cdots$ & $\cdots$ & $\cdots$ & $\cdots$                                                                                                                                                                \\ \cline{2-4} \cline{7-11}
\end{tabular}
  \caption{Additional new holographic quasiprimary operators in $\CFT^n$ with $W(2,4)$ symmetry. The notations here are the same as the ones in Table~\ref{quasi23ntab}.}\label{quasi24ntab}
\end{table}

\subsection{Calculation of coefficients}

The holomorphic quasiprimary operators that contribute to the one-loop mutual information are the ones in classes $WW$, $\mU\mU$, $TW\mU$, $TTWW$, $TW\mV$, $\mV\mV$, $TW\mX$. For operators in classes $WW$, $\mU\mU$ and $\mV\mV$ we have
\be
I_{WW}=I_{\mO\mO}|_{h=4}, ~~~ I_{\mU\mU}=I_{\mO\mO}|_{h=6}, ~~~ I_{\mV\mV}=I_{\mO\mO}|_{h=7}.
\ee

For operators in class $TW\mU$, we have
\bea
&& TW\mU, ~~~ I_1(TW\mU)=\ii\p TW\mU-\f12 T\ii\p W\mU, ~~~ I_2(TW\mU)=\ii\p T W\mU-\f13 TW\ii\p \mU, \nn\\
&& \II_1(TW\mU)=\p T\p W\mU-\f45\p^2TW\mU-\f29T\p^2W\mU,   \\
&& \II_2(TW\mU)=\p TW\p \mU-\f65\p^2TW\mU-\f2{13}TW\p^2\mU,   \nn\\
&& \II_3(TW\mU)=T\p W \p\mU-\f23 T\p^2W\mU-\f4{13}TW\p^2\mU,   \nn
\eea
the modified normalization factors
\be
\b_{TW\mU} = 1, ~~~
\b_{\sI(TW\mU)} = \frac{2}{3} \left(\begin{array}{cc} 9 & 6 \\ 6 & 8\end{array}\right), ~~~
\b_{\sII(TW\mU)} = \frac{16}{585} \left(\begin{array}{ccc} 2366 & 1404 & 780 \\ 1404 & 4131 & 540 \\ 780 & 540 & 6930\end{array}\right),
\ee
and the modified OPE coefficients
\bea
&& \hat b_{TW\mU}^{j_1j_2j_3} = \f{1}{2^{12}}\f{1}{s_{j_1j_3}^4 s_{j_2j_3}^8},                                                    ~~~
   \hat b_{\sI_1(TW\mU)}^{j_1j_2j_3} = \f{1}{2^{11}} \f{s_{j_1j_2}}{s_{j_1j_3}^5 s_{j_2j_3}^9},                                  \nn\\
&& \hat b_{\sI_2(TW\mU)}^{j_1j_2j_3} = \f{1}{3\cdot2^{11}} \f{s_{j_1j_2}-3s_{j_1j_2j_3}}{s_{j_1j_3}^5 s_{j_2j_3}^9},             ~~~
   \hat b_{\sII_1(TW\mU)}^{j_1j_2j_3} = \f{1}{2^{10}} \f{s_{j_1j_2}^2}{s_{j_1j_3}^6 s_{j_2j_3}^{10}},                               \nn\\
&& \hat b_{\sII_2(TW\mU)}^{j_1j_2j_3} = \f{1}{13\cdot2^{12}} \f{10s_{j_1j_2}^2-52s_{j_1j_3}^2+65s_{j_2j_3}^2+78s_{j_1j_2j_3}^2}
                                                               {s_{j_1j_3}^6 s_{j_2j_3}^{10}},                                      \\
&& \hat b_{\sII_3(TW\mU)}^{j_1j_2j_3} = \f{1}{13\cdot2^{11}} \f{36s_{j_1j_2}^2+117s_{j_1j_3}^2-104s_{j_2j_3}^2+78s_{j_1j_2j_3}^2}
                                                                {s_{j_1j_3}^6 s_{j_2j_3}^{10}}.                                      \nn
\eea
In class $TW\mV$, we have operators
\be
TW\mV, ~~~ \I_1(TW\mV)=\ii\p T W\mV-\f12 T\ii\p W\mV, ~~~ \I_2(TW\mV)=\ii\p T W\mV-\f27 TW\ii\p \mV,
\ee
the modified normalization factors
\be
\b_{TW\mV}=1, ~~~ \b_{\sI(TW\mV)}=\frac{2}{7} \left(\begin{array}{cc} 21 & 14 \\ 14 & 18\end{array}\right),
\ee
and the modified OPE coefficients
\bea
&& \hat b_{TW\mV}^{j_1j_2j_3} = -\f{\ii}{2^{13}}\f{s_{j_1j_2}}{s_{j_1j_3}^5 s_{j_2j_3}^9}, ~~~
   \hat b_{\sI_1(TW\mV)}^{j_1j_2j_3} = -\f{\ii}{2^{15}} \f{8s_{j_1j_2}^2 +s_{j_1j_3}^2 +2s_{j_2j_3}^2}
                                                          {s_{j_1j_3}^6 s_{j_2j_3}^{10}},                      \nn\\
&& \hat b_{\sI_2(TW\mV)}^{j_1j_2j_3} = -\f{\ii}{7\cdot2^{14}} \f{10s_{j_1j_2}^2 -28s_{j_1j_3}^2 +35s_{j_2j_3}^2}
                                                                {s_{j_1j_3}^6 s_{j_2j_3}^{10}}.
\eea
For operators in class $TW\mX$, we have
\be
TW\mX, ~~~ \b_{TW\mX}=1, ~~~ \hat b_{TW\mX}^{j_1j_2j_3}=\f{1}{2^{14}}\f{s_{j_1j_2}^2}{s_{j_1j_3}^6 s_{j_2j_3}^{10}}.
\ee

In class $TTWW$, we choose $C_K=\f{c^2}{8}$ and we have operators
\bea
&& TTWW, ~~~ \I_1(TTWW)=\ii\p TTWW-T\ii\p TWW,                                            \nn\\
&& \I_2(TTWW)=\ii\p TTWW- \f12 TT\ii\p WW, ~~~ \I_3(TTWW)=\ii\p TTWW - \f12 TTW\ii\p W,    \nn\\
&& \II_1(TTWW)=\p T \p TWW - \f25 \p^2 TTWW - \f25 T \p^2TWW,                             \nn\\
&& \II_2(TTWW)=\p T T\p WW - \f45 \p^2 TTWW - \f29 TT\p^2WW,                             \nn\\
&& \II_3(TTWW)=\p T TW\p W - \f45 \p^2 TTWW - \f29 TTW\p^2W,                             \\
&& \II_4(TTWW)=T \p T\p WW - \f45 T\p^2 TWW - \f29 TT\p^2WW,                             \nn\\
&& \II_5(TTWW)= T \p TW\p W - \f45 T\p^2 TWW - \f29 TTW\p^2W,                             \nn\\
&& \II_6(TTWW)=T T\p W\p W - \f49 TT\p^2 WW - \f49 TTW\p^2W,                             \nn
\eea
modified normalization factors
\be
\b_{TTWW}=1, ~~~
\b_{\sI(TTWW)}=2 \left(\begin{array}{ccc} 4 & 2 & 2 \\ 2 & 3 & 2 \\ 2 & 2 & 3\end{array}\right), ~~~
\b_{\sII(TTWW)}=\frac{16}{45} \left(
\begin{array}{cccccc}
 81 & 36 & 36 & 36 & 36 & 0 \\
 36 & 182 & 72 & 20 & 0 & 40 \\
 36 & 72 & 182 & 0 & 20 & 40 \\
 36 & 20 & 0 & 182 & 72 & 40 \\
 36 & 0 & 20 & 72 & 182 & 40 \\
 0 & 40 & 40 & 40 & 40 & 340
\end{array}
\right),
\ee
and modified OPE coefficients
\bea
&& \hat b_{TTWW}^{j_1j_2j_3j_4}=\f{1}{2^{12}}\f{1}{s_{j_1j_2}^4s_{j_3j_4}^8},                           ~~~
   \hat b_{\sI_1(TTWW)}^{j_1j_2j_3j_4}=-\f{1}{2^{10}}\f{c_{j_1j_2}}{s_{j_1j_2}^5s_{j_3j_4}^8},             \nn\\
&& \hat b_{\sI_2(TTWW)}^{j_1j_2j_3j_4}=\f{1}{2^{11}}\f{s_{j_1j_2j_3j_4}}{s_{j_1j_2}^5s_{j_3j_4}^9},      ~~~
   \hat b_{\sI_3(TTWW)}^{j_1j_2j_3j_4}=\hat b_{\sI_2(TTWW)}^{j_1j_2j_4j_3},                              \nn\\
&& \hat b_{\sII_1(TTWW)}^{j_1j_2j_3j_4}=\f{1}{2^{12}}\f{9-8s_{j_1j_2}^2}{s_{j_1j_2}^6s_{j_3j_4}^8},     ~~~
   \hat b_{\sII_2(TTWW)}^{j_1j_2j_3j_4}=\f{1}{2^{10}}\f{s_{j_1j_2j_3j_4}^2}{s_{j_1j_2}^6s_{j_3j_4}^{10}},  \\
&& \hat b_{\sII_3(TTWW)}^{j_1j_2j_3j_4} = \hat b_{\sII_2(TTWW)}^{j_1j_2j_4j_3},                          ~~~
   \hat b_{\sII_4(TTWW)}^{j_1j_2j_3j_4} = \hat b_{\sII_2(TTWW)}^{j_2j_1j_3j_4},                          \nn\\
&& \hat b_{\sII_5(TTWW)}^{j_1j_2j_3j_4} = \hat b_{\sII_2(TTWW)}^{j_2j_1j_4j_3},                          ~~~
   \hat b_{\sII_6(TTWW)}^{j_1j_2j_3j_4}=\f{1}{2^{11}}\f{17-16s_{j_3j_4}^2}{s_{j_1j_2}^4s_{j_3j_4}^{10}}.   \nn
\eea

\subsection{One-loop mutual information}

The contributions from operators of different classes are respectively
\bea
&& \hspace{-5mm}
   I_{WW}^{\oloop} = \frac{x^8}{218790}+\frac{4 x^9}{230945}+\frac{3 x^{10}}{76076}+\frac{5 x^{11}}{71162}+\frac{11 x^{12}}{101660}+\frac{11 x^{13}}{72675}+\frac{1001 x^{14}}{5058180}+O(x^{15}),                                                             \nn\\
&& \hspace{-5mm}
   I_{\mU\mU}^{\oloop} = \frac{x^{12}}{67603900}+\frac{x^{13}}{11700675}+\frac{13 x^{14}}{46535256}+O(x^{15}),       ~~~
   I_{TW\mU}^\oloop = -\frac{x^{12}}{33801950}-\frac{2 x^{13}}{11700675}-\frac{13 x^{14}}{23267628}+O(x^{15}),       \nn\\
&& \hspace{-5mm}
   I_{TTWW}^\oloop = \frac{x^{12}}{67603900}+\frac{x^{13}}{11700675}+\frac{163 x^{14}}{581690700}+O(x^{15}),         ~~~
   I_{TW\mV}^\oloop = -\frac{x^{14}}{581690700}+O\left(x^{15}\right),                                                \\
&& \hspace{-5mm}
   I_{\mV\mV}^\oloop = \frac{x^{14}}{1163381400}+O(x^{15}),                                                          ~~~
   I_{TW\mX}^\oloop = O(x^{15}).                                                                                     \nn
\eea
Summing them together, we get the additional contributions of $W_4$ operator to one-loop mutual information
\be
I_{(4)}^\oloop = \frac{x^8}{218790}+\frac{4 x^9}{230945}+\frac{3 x^{10}}{76076}+\frac{5 x^{11}}{71162}+\frac{11 x^{12}}{101660}+\frac{11 x^{13}}{72675}+\frac{1001 x^{14}}{5058180}+O(x^{15}),
\ee
and this matches the gravity result in \cite{Beccaria:2014lqa}, i.e. $I_{\spin 4}^\oloop$ in (\ref{mutual234}). Note that there is cancellation
\be \label{cancel2}
I_{\mU\mU}^{\oloop}+I_{TW\mU}^{\oloop}+I_{TTWW}^{\oloop}+I_{TW\mV}^{\oloop}+I_{\mV\mV}^{\oloop}+I_{TW\mX}^{\oloop}=O(x^{15}).
\ee

\section{Conclusion and discussion}\label{s6}

In this paper we have calculated the one-loop entanglement entropy of  two short intervals using OPE of twist operators in the CFT side. Following the strategy in \cite{Beccaria:2014lqa} we took the $n\to1$ limit of the R\'enyi entropy, and this allows us to get the one-loop entanglement entropy with higher order of the cross ratio $x$ than before. We considered the contributions of stress tensor, $W_3$ operator, and $W_4$ operator. The results are in agreement with the ones of gravity side in \cite{Beccaria:2014lqa}. It is notable that there are nontrivial cancellations in (\ref{cancel1}) and (\ref{cancel2}). We do not know if there may be some further indications for these cancellations.

In the gravity side, contributions of general spin-$s$ fields to the entanglement entropy have been organized into different parts \cite{Beccaria:2014lqa}. It would be nice to investigate if one can organize the CFT$^n$ quasiprimary operators that appear in the OPE of twist operators so that some particular quasiprimary operators contribute to some particular parts of the entanglement entropy. It is expected that there are cancellations similar to (\ref{cancel1}) and (\ref{cancel2}) in contributions of a $W_s$ operator with general $s$ to the one-loop entanglement entropy.

\section*{Acknowledgments}

We would like to thank Hai Lin, Wei Song, Qiang Wen, and Jun-Bao Wu for helpful discussions.
We thank Matthew Headrick for his Mathematica code \emph{Virasoro.nb} that could be downloaded at \url{http://people.brandeis.edu/~headrick/Mathematica/index.html}.
The work was in part supported by NSFC Grants No.~11222549 and No.~11575202.

\appendix

\section{Contributions of new quasiprimary operators with two old ones}\label{classoo}

In this appendix, we investigate the contributions of new holomorphic quasiprimary operators of $\CFT^n$ with two old holomorphic quasiprimary operators to the one-loop mutual information. We consider a general old holomorphic quasiprimary $\mO$ with an integer conformal dimension $(h,0)$. Using two of them we construct the new quasiprimary operators to order $2h+6,$\footnote{Some of the operators have been constructed in \cite{Chen:2014kja,Beccaria:2014lqa}, and the corresponding coefficients $\a_K$ and $d_K$ have also been calculated therein.}
\bea
&& \mO\mO, ~~~ \I(\mO\mO)=\mO\ii\p\mO-\ii\p\mO\mO, ~~~ \II(\mO\mO)=\p\mO\p\mO-\f{h}{2h+1}(\mO\p^2\mO+\p^2\mO\mO),                 \nn\\
&& \III(\mO\mO)=\ii\p\mO\p^2\mO-\p^2\mO\p\ii\mO-\f{h}{3(h+1)}( \mO\ii\p^3\mO - \ii\p^3\mO\mO ),                                   \nn\\
&& \IV(\mO\mO)=\p^2\mO\p^2\mO - \f{2h+1}{3(h+1)}(\p\mO\p^3\mO+\p^3\mO\p\mO) + \f{h(2h+1)}{6(h+1)(2h+3)}(\mO\p^4\mO+\p^4\mO\mO),   \nn\\
&& \V(\mO\mO)= \p^2\mO\ii\p^3\mO-\ii\p^3\mO\p^2\mO-\f{2h+1}{2(2h+3)}( \ii\p\mO\p^4\mO -\p^4\mO\ii\p\mO )                          \\
&& \phantom{\V(\mO\mO)=}
               +\f{h(2h+1)}{10(h+2)(2h+3)}( \mO\ii\p^5\mO - \ii\p^5\mO\mO ),                                                       \nn\\
&& \VI(\mO\mO) = \p^3\mO\p^3\mO-\f{3(h+1)}{2(2h+3)}(\p^2\mO\p^4\mO+\p^4\mO\p^2\mO)
                +\f{3(h+1)(2h+1)}{10(h+2)(2h+3)}(\p\mO\p^5\mO+\p^5\mO\p\mO) \nn\\
&& \phantom{\VI(\mO\mO)=}
                -\f{h(h+1)(2h+1)}{10(h+2)(2h+3)(2h+5)}(\mO\p^6\mO+\p^6\mO\mO).                                               \nn
\eea
Note that we have omitted the subscripts $j_1,j_2=0,1,\cdots,n-1$ with $j_1<j_2$, and so each equation above actually represent $\f{n(n-1)}{2}$ operators.

The normalization of $\mO$ is $\a_\mO$, and for all these operators we choose $C_K=\a_\mO$ and $\td\a_K=\a_\mO^2$. Then we get the modified normalization factor
\bea
&& \b_{\mO\mO}=1, ~~~ \b_{\sI(\mO\mO)}=4h, ~~~ \b_{\sII(\mO\mO)}=\frac{4 h^2 (4 h+1)}{2 h+1},\nn\\
&& \b_{\sIII(\mO\mO)}=\frac{16 h^2 (2 h+1) (4 h+3)}{3 (h+1)},     \nn\\
&& \b_{\sIV(\mO\mO)}=\frac{16 h^2 (2 h+1)^2 (4 h+3) (4 h+5)}{3 (h+1) (2 h+3)},  \\
&& \b_{\sV(\mO\mO)}=\frac{192 h^2 (h+1) (2 h+1)^2 (4 h+5) (4 h+7)}{5 (h+2) (2 h+3)},  \nn\\
&& \b_{\sVI(\mO\mO)}=\frac{576 h^2 (h+1)^2 (2 h+1)^2 (4 h+5) (4 h+7) (4 h+9)}{5 (h+2) (2 h+3) (2 h+5)}. \nn
\eea
We also have the modified OPE coefficients
\bea
&& \hspace{-5mm}
   \hat b_{\mO\mO}^{j_1j_2}=\f{1}{(2\ii)^{2h}}\f{1}{s_{j_1j_2}^{2h}}, ~~~
   \hat b_{\sI(\mO\mO)}^{j_1j_2}=\f{2h}{(2\ii)^{2h}}\f{c_{j_1j_2}}{s_{j_1j_2}^{2h+1}}, ~~~
   \hat b_{\sII(\mO\mO)}^{j_1j_2} = \f{h}{2(2\ii)^{2h}}\f{(4h+1)-4h s_{j_1j_2}^2}{s_{j_1j_2}^{2h+2}}, \nn\\
&& \hspace{-5mm}
   \hat b_{\sIII(\mO\mO)}^{j_1j_2} = \f{h(2h+1)}{3(2\ii)^{2h}}\f{c_{j_1j_2}\big[ (4h+3)-4hs_{j_1j_2}^2 \big]}{s_{j_1j_2}^{2h+3}},   \nn\\
&& \hspace{-5mm}
   \hat b_{\sIV(\mO\mO)}^{j_1j_2} = \f{h(2h+1)}{12(2\ii)^{2h}}
                                    \f{(4h+3)(4h+5)-4 (2h+1)(4h+3) s_{j_1j_2}^2+8h(2h+1)s_{j_1j_2}^4}{s_{j_1j_2}^{2h+4}} ,  \\
&& \hspace{-5mm}
   \hat b_{\sV(\mO\mO)}^{j_1j_2} = \f{h(h+1)(2h+1)}{10(2\ii)^{2h}}
                 \f{c_{j_1j_2}\big[ (4h+5)(4h+7)-4 (2h+1)(4h+5) s_{j_1j_2}^2+8h(2h+1)s_{j_1j_2}^4 \big]}{s_{j_1j_2}^{2h+5}},  \nn\\
&& \hspace{-5mm}
   \hat b_{\sVI(\mO\mO)}^{j_1j_2} = \f{h(h+1)(2h+1)}{40(2\ii)^{2h}s_{j_1j_2}^{2h+6}}
                                   \big[ (4h+5)(4h+7)(4h+9)-12(h+1)(4h+5)(4h+7)s_{j_1j_2}^2           \nn\\
&& \hspace{-5mm}
   \phantom{\hat b_{\sVI(\mO\mO)}^{j_1j_2} = \f{h(h+1)(2h+1)}{40(2\ii)^{2h}s_{j_1j_2}^{2h+6}}\big[}
                                        +24(h+1)(2h+1)(4h+5)s_{j_1j_2}^4-32h(h+1)(2h+1)s_{j_1j_2}^6 \big],  \nn
\eea
with the definitions $s_{j_1j_2}=\sin(\f{j_1-j_2}{n}\pi)$ and $c_{j_1j_2}=\cos(\f{j_1-j_2}{n}\pi)$.
Using (\ref{fm}), (\ref{fmne1}) and taking into the contributions of the antiholomorphic sector, we get the contributions of the above operators to the mutual information
\bea \label{IOO}
&& \hspace{-5mm}
   I_{\mO\mO}=\f{\G(3/2)\G(2h+1)}{\G(2h+3/2)}\Big(\f{x}{4}\Big)^{2h} \Big[
   1+\frac{2 h (2 h+1) x}{4 h+3}+\frac{(h+1) (2 h+1)^2 (4 h+1) x^2}{2 (16 h^2+32 h+15)} \nn\\
&& \hspace{10mm}
   +\frac{(h+1)^2 (2 h+1)^2 (2 h+3) x^3}{3 (16 h^2+48 h+35)}+\frac{(h+1)^2 (h+2) (2 h+1) (2 h+3)^2 x^4}{12 (16 h^2+64 h+63)} \nn\\
&& \hspace{10mm}
  +\frac{(h+1)^2 (h+2)^2 (2 h+1) (2 h+3)^2 (2 h+5) x^5}{30 (64 h^3+368 h^2+636 h+297)} \\
&& \hspace{10mm}
  +\frac{(h+1) (h+2)^2 (h+3) (2 h+1) (2 h+3)^2 (2 h+5)^2 x^6}{360 (64 h^3+432 h^2+860 h+429)}+O(x^7) \Big]. \nn
\eea
Note that these operators only contribute to the one-loop part of the mutual information.
In \cite{Beccaria:2014lqa} there is the gravity result that for spin-$s$ field one part of the one-loop entanglement entropy is
\be
S_{\textrm{CDW},(\sI)}^{(s)\oloop}=-\f{\G(3/2)\G(2h+1)}{\G(2h+3/2)}\Big(\f{x}{4}\Big)^{2s}{}_3F_2(2s,2s-1/2,2s+1;2s+3/2,4s-1;x),
\ee
and our result (\ref{IOO}) is in accord with this by setting $h=s$.

\section{Some summation formulas}\label{sum}

We collect some useful summation formulas in this appendix. Firstly we define
\be \label{fm}
f_m=\sum_{j=1}^{n-1}\f{1}{ \lt( \sin\f{\pi j}{n} \rt)^{2m}},
\ee
with $m$ being an integer. As shown in \cite{Calabrese:2010he}, one has $f_m\sim n-1$ and
\be \label{fmne1}
\td f_m=\lim_{n \to 1}\f{f_m}{n-1}=\f{\Gamma ({3}/{2})\Gamma (m+1)}{\Gamma (m+{3}/{2})}.
\ee
There are several summations that are related to (\ref{fm}), and these include
\bea
&& \sum_{j_1,j_2}^{\neq} \f{1}{s_{j_1j_2}^{2m}}=n f_m, ~~~
   \sum_{j_1,j_2,j_3}^{\neq} \f{1}{s_{j_1j_2}^{2p}s_{j_1j_3}^{2q}} = n(f_p f_q - f_{p+q}), ~~~                    \nn\\
&& \sum_{j_1,j_2,j_3}^{\neq} \f{c_{j_1j_2}c_{j_1j_3}}{s_{j_1j_2}^{2p+1}s_{j_1j_3}^{2q+1}} = n(f_{p+q}-f_{p+q+1}), ~~~\\
&& \sum_{j_1,j_2,j_3,j_4}^{\neq} \f{1}{s_{j_1j_2}^{2p}s_{j_3j_4}^{2q}} = n(n-4)f_p f_q + 2n f_{p+q}.\nn
\eea
All the summations indices in above equations are in the range $0 \leq j_{1,2,3,4} \leq n-1$. The first summation has the constraint $j_1\neq j_2$, the second and third summations have the constraints $j_1\neq j_2$, $j_1\neq j_3$, and $j_2 \neq j_3$, and the last summation has the constraints $j_1\neq j_2$, $j_1\neq j_3$, $j_1 \neq j_4$, $j_2\neq j_3$, $j_2\neq j_4$, and $j_3 \neq j_4$. We use the same summation notations below.

We define that
\bea
&& s_{p,q} = \sum_{j_1,j_2,j_3}^{\neq} \f{s_{j_1j_2}^2}{s_{j_1j_3}^{2p}s_{j_2j_3}^{2q}}, ~~~
   t_{p,q} = \sum_{j_1,j_2,j_3}^{\neq} \f{s_{j_1j_2}^4}{s_{j_1j_3}^{2p}s_{j_2j_3}^{2q}}, \nn\\
&& u_{p,q} = \sum_{j_1,j_2,j_3}^{\neq} \f{s_{j_1j_2}^2c_{j_1j_3}c_{j_2j_3}}{s_{j_1j_3}^{2p+1}s_{j_2j_3}^{2q+1}}, ~~~
   v_{p,q} = \sum_{j_1,j_2,j_3}^{\neq} \f{s_{j_1j_2}c_{j_1j_3}}{s_{j_1j_3}^{2p}s_{j_2j_3}^{2q+1}}.
\eea
We have
\bea
&& \hspace{-5mm} s_{5,5}=\frac{4n (n^2-1)^2 (n^2-4) (n^2+11) (3 n^4+10 n^2+227) (5 n^6+58 n^4+325 n^2+1052)}{6630710625},  \nn\\
&& \hspace{-5mm} s_{5,6}= \frac{2 n(n^2-1)^2(n^2-4)}{99560120034375}(45553 n^{14}+1328108 n^{12}+19669231 n^{10}+201786116 n^8+1535925879 n^6 \nn\\
&& \hspace{-5mm} \phantom{s_{5,6}=} +8192615444 n^4+29746589337 n^2+64811480332),\nn\\
&& \hspace{-5mm} s_{5,7}= \frac{2 n(n^2-1)^2(n^2-4)}{298680360103125}(13840 n^{16}+448758 n^{14}+7377133 n^{12}+83185441 n^{10}+691628526 n^8 \nn\\
&& \hspace{-5mm} \phantom{s_{5,7}=} +4541914744 n^6+22337114089 n^4+77828433057 n^2+166234428412),\nn\\
&& \hspace{-5mm} s_{5,8}= \frac{n(n^2-1)^2(n^2-4)}{76163491826296875}(715083 n^{18}+25534538 n^{16}+461805573 n^{14}+5699916578 n^{12}+52320956483 n^{10} \nn\\
&& \hspace{-5mm} \phantom{s_{5,8}=} +375059702238 n^8+2232468198983 n^6+10411222626638 n^4+35326719643878 n^2+74499122340008) ,\nn\\
&& \hspace{-5mm} s_{5,9}= \frac{2 n(n^2-1)^2(n^2-4)}{30389233238692453125} (14453970 n^{20}+563655376 n^{18}+11124486091 n^{16}+149347794891 n^{14} \nn\\
&& \hspace{-5mm} \phantom{s_{5,9}=} +1500571528631 n^{12}+11858181395071 n^{10}+76527402573861 n^8+425332156697681 n^6 \nn\\
&& \hspace{-5mm} \phantom{s_{5,9}=} +1912070720866171 n^4+6372177472656981 n^2+13322930703091276),\nn\\
&& \hspace{-5mm} s_{5,10}= \frac{2 n(n^2-1)^2(n^2-4)}{455838498580386796875} (21967243 n^{22}+928905348 n^{20}+19867081060 n^{18}+288349232835 n^{16} \nn\\
&& \hspace{-5mm} \phantom{s_{5,10}=} +3144899150355 n^{14}+27151604038455 n^{12}+192213812991645 n^{10}+1148722121535645 n^8 \\
&& \hspace{-5mm} \phantom{s_{5,10}=} +6073947641495190 n^6+26593077984745265 n^4+87503665266114507 n^2+181852852200342452),\nn\\
&& \hspace{-5mm} s_{6,6}= \frac{4 n(n^2-1)^2(n^2-4) (2 n^8+35 n^6+321 n^4+2125 n^2+14797)}{59736072020625}(691 n^8+10280 n^6+75663 n^4 \nn\\
&& \hspace{-5mm} \phantom{s_{6,6}=} +355070 n^2+1070296),\nn\\
&& \hspace{-5mm} s_{6,7}= \frac{2 n(n^2-1)^2(n^2-4)}{407698691540765625} (1910462 n^{18}+68101172 n^{16}+1226741277 n^{14}+14905903687 n^{12}+139242875522 n^{10} \nn\\
&& \hspace{-5mm} \phantom{s_{6,7}=} +1046414082282 n^8+6136429840777 n^6+27331736137187 n^4+89096568481962 n^2+183491159525672),\nn\\
&& \hspace{-5mm} s_{6,9}= \frac{2 n(n^2-1)^2(n^2-4)}{5925900481545028359375} (285030529 n^{22}+12033503724 n^{20}+256501410985 n^{18}+3681748010085 n^{16} \nn\\
&& \hspace{-5mm} \phantom{s_{6,9}=} +40260080243145 n^{14}+353054399664045 n^{12}+2550512661662865 n^{10}+15681395452923615 n^8 \nn\\
&& \hspace{-5mm} \phantom{s_{6,9}=} +80490367507898250 n^6+331149869650451675 n^4+1032349661011754226 n^2+2076990250139446856),\nn\\
&& \hspace{-5mm} s_{6,10}= \frac{2 n(n^2-1)^2(n^2-4)}{1368883011236901551015625} (6671146880 n^{24}+303588222092 n^{22}+6973032426942 n^{20} \nn\\
&& \hspace{-5mm} \phantom{s_{6,10}=} +107805919807535 n^{18}+1267311092051085 n^{16}+11977039037640765 n^{14}+93701187618319965 n^{12} \nn\\
&& \hspace{-5mm} \phantom{s_{6,10}=} +619839214947433575 n^{10}+3572385606811692975 n^8+17593549852111719955 n^6 \nn\\
&& \hspace{-5mm} \phantom{s_{6,10}=} +70678251505956283825 n^4+217481207149114176078 n^2+434593017048771078328), \nn
\eea
\bea
&& \hspace{-5mm} t_{6,6}= \frac{4 n(n^2-1)^2(n^2-4)}{99560120034375} (91053 n^{14}+1751258 n^{12}+15802641 n^{10}+86621976 n^8+122885159 n^6 \nn\\
&& \hspace{-5mm} \phantom{t_{6,6}=} -1216809126 n^4-8233724853 n^2-22129450108),\nn\\
&& \hspace{-5mm} t_{6,7}= \frac{2 n(n^2-1)^2(n^2-4)}{298680360103125}(55300 n^{16}+1244154 n^{14}+13403599 n^{12}+89917783 n^{10}+353556798 n^8 \nn\\
&& \hspace{-5mm} \phantom{t_{6,7}=} +63068872 n^6-7097114693 n^4-39444148809 n^2-100293199004),\nn\\
&& \hspace{-5mm} t_{6,8}= \frac{n(n^2-1)^2(n^2-4)}{6930877756193015625} (259976261 n^{18}+6702162006 n^{16}+83850278121 n^{14}+668341854016 n^{12} \nn\\
&& \hspace{-5mm} \phantom{t_{6,8}=} +3453872052321 n^{10}+10488470468166 n^8-12276550624049 n^6-301684860616404 n^4  \\
&& \hspace{-5mm} \phantom{t_{6,8}=} -1511339984382654 n^2-3720039665967784),\nn\\
&& \hspace{-5mm} t_{6,10}= \frac{2 n(n^2-1)^2(n^2-4)}{5925900481545028359375} (1140875688 n^{22}+36914466948 n^{20}+588438944965 n^{18}+6119943169290 n^{16} \nn\\
&& \hspace{-5mm} \phantom{t_{6,10}=} +45084693422310 n^{14}+239240140814910 n^{12}+866020397951400 n^{10}+1596578940865050 n^8 \nn\\
&& \hspace{-5mm} \phantom{t_{6,10}=} -9709837053108030 n^6-105197092471166730 n^4-475582750715286333 n^2-1131502803657349468), \nn
\eea
\bea
&& \hspace{-5mm} u_{5,5}= -\frac{8 n(n^2-1)^2(n^2-4)^2 (5 n^6+58 n^4+325 n^2+1052)^2}{218813450625},\nn\\
&& \hspace{-5mm} u_{5,6}= -\frac{8 n(n^2-1)^2(n^2-4)^2 (5 n^6+58 n^4+325 n^2+1052) (691 n^8+10280 n^6+75663 n^4+355070 n^2+1070296)}{298680360103125},\nn\\
&& \hspace{-5mm} u_{5,7}= -\frac{16 n(n^2-1)^2(n^2-4)^2 (5 n^6+58 n^4+325 n^2+1052)}{896041080309375}(105 n^{10}+1907 n^8+17305 n^6+102921 n^4\\
&& \hspace{-5mm} \phantom{u_{5,7}=} +436090 n^2+1256072),\nn\\
&& \hspace{-5mm} u_{5,9}= -\frac{8 n(n^2-1)^2(n^2-4)^2 (5 n^6+58 n^4+325 n^2+1052)}{91167699716077359375}(219335 n^{14}+5426224 n^{12}+67562250 n^{10}\nn\\
&& \hspace{-5mm} \phantom{u_{5,9}=} +561172268 n^8+3465459895 n^6+16695492492 n^4+63127741520 n^2+172125054016),\nn
\eea
\bea
&&\hspace{-5mm} v_{5,6} = -\frac{2 n(n^2-1)^2(n^2-4) (5 n^6+58 n^4+325 n^2+1052)}{298680360103125}(1382 n^{10}+28682 n^8+307961 n^6+2295661 n^4\nn\\
&& \hspace{-5mm} \phantom{v_{5,6}=} +13803157 n^2+92427157), \nn\\
&&\hspace{-5mm} v_{5,8} = -\frac{2 n(n^2-1)^2(n^2-4) (5 n^6+58 n^4+325 n^2+1052)}{228490475478890625}(10851 n^{14}+296451 n^{12}+4149467 n^{10}\nn\\
&& \hspace{-5mm} \phantom{v_{5,6}=} +39686267 n^8+292184513 n^6+1777658113 n^4+9611679169 n^2+61430943169), \\
&&\hspace{-5mm} v_{7,4} = -\frac{4 n(n^2-1)^2(n^2-4) (n^2+11) (3 n^4+10 n^2+227)}{27152760009375}(105 n^{10}+1907 n^8+17305 n^6+102921 n^4\nn\\
&& \hspace{-5mm} \phantom{v_{5,6}=} +436090 n^2+1256072), \nn\\
&&\hspace{-5mm} v_{9,4} = -\frac{2 n(n^2-1)^2(n^2-4) (n^2+11) (3 n^4+10 n^2+227)}{2762657567153859375}(219335 n^{14}+5426224 n^{12}+67562250 n^{10}\nn\\
&& \hspace{-5mm} \phantom{v_{5,6}=} +561172268 n^8+3465459895 n^6+16695492492 n^4+63127741520 n^2+172125054016).\nn
\eea

We define
\bea
&& a_{p,q,r,s}=\sum_{j_1,j_2,j_3,j_4}^{\neq}\f{1}{s_{j_1j_2}^{2p}s_{j_3j_4}^{2q}s_{j_1j_3}^{2r}s_{j_2j_4}^{2s}} ,  \nn\\
&& b_{p,q,r,s}=\sum_{j_1,j_2,j_3,j_4}^{\neq}\f{c_{j_1j_2}c_{j_1j_3}}
                                              {s_{j_1j_2}^{2p+1}s_{j_3j_4}^{2q}s_{j_1j_3}^{2r+1}s_{j_2j_4}^{2s}} ,  \\
&& c_{p,q,r,s}=\sum_{j_1,j_2,j_3,j_4}^{\neq}\f{c_{j_1j_2}c_{j_3j_4}c_{j_1j_3}c_{j_2j_4}}
                                              {s_{j_1j_2}^{2p+1}s_{j_3j_4}^{2q+1}s_{j_1j_3}^{2r+1}s_{j_2j_4}^{2s+1}} ,  \nn\\
&& d_{p,q,r,s}=\sum_{j_1,j_2,j_3,j_4}^{\neq}\f{c_{j_1j_2}c_{j_3j_4}}
                                              {s_{j_1j_2}^{2p+1}s_{j_3j_4}^{2q+1}s_{j_1j_3}^{2r}s_{j_2j_4}^{2s}}. \nn
\eea
We have
\bea
&& \hspace{-5mm} a_{2,2,2,2}= \frac{4 n(n^2-1)(n^2-4)(n^2-9)}{54273594375}(21 n^{10}+1994 n^8+105648 n^6+4785522 n^4+141534331 n^2+2127620484), \nn\\
&& \hspace{-5mm} a_{2,2,2,3}= \frac{2 n(n^2-1)(n^2-4)(n^2-9)}{194896477400625} (11120 n^{12}+1096256 n^{10}+59129609 n^8+2551249273 n^6+80209669687 n^4\nn\\
&& \hspace{-5mm}\phantom{a_{2,2,2,3}=} +1740731207971 n^2+22574176404084), \nn\\
&& \hspace{-5mm} a_{2,3,2,3}= \frac{2 n(n^2-1)(n^2-4)(n^2-9)}{32157918771103125} (118663 n^{14}+12334362 n^{12}+688418766 n^{10}+28724959384 n^8\\
&& \hspace{-5mm}\phantom{a_{2,3,2,3}=}+988368726279 n^6+26290949481846 n^4+506319939369292 n^2+6044979290991408), \nn\\
&& \hspace{-5mm} a_{3,3,3,3}= \frac{2 n(n^2-1)(n^2-4)(n^2-9)}{3028793579456347828125}(23125517 n^{18}+3014718238 n^{16}+205212344463 n^{14}+9781329558932 n^{12}\nn\\
&& \hspace{-5mm}\phantom{a_{3,3,3,3}=} +371907750192979 n^{10}+12650688547612206 n^8+368606252765781785 n^6+9113476333879452640 n^4\nn\\
&& \hspace{-5mm}\phantom{a_{3,3,3,3}=} +174930349600869335256 n^2+2059147054618331397984), \nn
\eea
\bea
&& \hspace{-5mm} b_{2,2,2,2} = -\frac{2 n(n^2-1)(n^2-4)(n^2-9)}{194896477400625}(3755 n^{12}+386267 n^{10}+21570773 n^8+929319721 n^6+32134979884 n^4 \nn\\
&& \hspace{-5mm} \phantom{b_{2,2,2,2}=}+854833946512 n^2+12005931953088),
\eea
\bea
&& \hspace{-5mm} c_{2,2,2,2} = -\frac{2 n(n^2-1)(n^2-4)(n^2-9)}{6431583754220625} (9401 n^{14}+529374 n^{12}+10085796 n^{10}-183846586 n^8-21275167485 n^6\nn\\
&& \hspace{-5mm} \phantom{c_{2,2,2,2}=}-1024689595860 n^4-31153950441712 n^2-468739784852928),
\eea
\bea
&& \hspace{-5mm} d_{2,2,2,2} = -\frac{2 n(n^2-1)(n^2-4)(n^2-9)}{64965492466875}(510 n^{12}+75236 n^{10}+5064199 n^8+233577843 n^6+9882627307 n^4\nn\\
&& \hspace{-5mm} \phantom{d_{2,2,2,2}=}+280124377421 n^2+3986367653484), \\
&& \hspace{-5mm} d_{2,2,2,3} = -\frac{2 n(n^2-1)(n^2-4)(n^2-9)}{1531329465290625}(283 n^{14}+84812 n^{12}+7204196 n^{10}+371271219 n^8+15346533334 n^6\nn\\
&& \hspace{-5mm} \phantom{d_{2,2,2,3}=}+519473950801 n^4+12716870657687 n^2+170109742777668). \nn
\eea

%\bibliographystyle{D:/00.zbib/utphys}   %%非常好, 期刊, arXiv超链接
%\bibliography{D:/00.zbib/1970,D:/00.zbib/1980,D:/00.zbib/1990,D:/00.zbib/1995,D:/00.zbib/1996,D:/00.zbib/1997,D:/00.zbib/1998,D:/00.zbib/1999,D:/00.zbib/2000,D:/00.zbib/2001,D:/00.zbib/2002,D:/00.zbib/2003,D:/00.zbib/2004,D:/00.zbib/2005,D:/00.zbib/2006,D:/00.zbib/2007,D:/00.zbib/2008,D:/00.zbib/2009,D:/00.zbib/2010,D:/00.zbib/2011,D:/00.zbib/2012,D:/00.zbib/2013,D:/00.zbib/2014,D:/00.zbib/2015,D:/00.zbib/2016,D:/00.zbib/book,D:/00.zbib/work,D:/00.zbib/thesis}

\providecommand{\href}[2]{#2}\begingroup\raggedright\endgroup

\end{document}